\documentclass[onecolumn,11pt,pre,longbibliography]{revtex4-1}
\usepackage{centernot}
\usepackage{mathtools}
\usepackage{stmaryrd}
\usepackage{amsmath,amssymb,amsfonts}
\usepackage{color}
\usepackage{graphicx}
\usepackage{subfig}
\usepackage{caption}
\usepackage{float}
\usepackage{comment}
\usepackage{multirow}

\begin{document}

\title{Post-blowup dynamics for the nonlinear Schr\"odinger equation}
\date{\today}
\author{José M. Escorcia}\email{jmescorcit@eafit.edu.co}
\affiliation{Universidad EAFIT, Carrera 49 No. 7 Sur-50, Medellín 050022, Colombia.}
\author{Alexei A. Mailybaev}\email{alexei@impa.br}
\affiliation{Instituto de Matem\'atica Pura e Aplicada -- IMPA, Rio de Janeiro, Brazil.}

\begin{abstract}

In this work we present a systematic numerical study of the post-blowup dynamics  of singular solutions of the 1D focusing critical NLS equation in the framework of a nonlinear damped perturbation. The first part of this study  shows that initially the post-blowup is  described by the adiabatic approximation, in which the collapsing core approaches an universal profile and the solution width  is governed by a system of  ODEs (reduced system). After that,  a non-adiabatic regime is observed soon after the maximum of the solution, in which our direct numerical simulations show a clear deviation from the dynamics based on the reduced system. Our study suggests that such non-adiabatic regime is caused by the increasing influx of mass into the collapsing core of the solution, which is not considered in the derivation of the reduced system.  Also, adiabatic theoretical predictions related to the wave-maximum and wave-dissipation  are compared with our numerical simulations. The second part of this work describes  the non-adiabatic dynamics. Here, numerical simulations reveal a dominant  quasi linear regime, caused by the rapid defocusing process.  The collapsing core approaches the universal profile, after removing some oscillations resulting from the interference with the tail. Finally, our numerical study suggests that in the limit of vanishing dissipation, and in a free-space domain, the critical mass is radiated to  infinity instantly at the collapse time.
\end{abstract}

\maketitle

\section{Introduction}

%%%%%%%%%%%%%%%%%%%%%%%%%%%%%%%%%%%%
 The nonlinear Schr\"odnger equation (NLS) 
 \begin{equation} 
 \label{eqNLS1}
    i\psi_t+  \Delta \psi + |\psi|^{2\sigma} \psi = 0,
    \end{equation}
 appears as fundamental model in different branches of physics, e.g., Bose-Einstein condensates \cite{Gerton00}, fluid dynamics \cite{Slunyaev13} and nonlinear optics \cite{Fibich15,Gabitov2002}. Here $\psi$ is a complex-valued function depending of the time $t$ and the space variable $\boldsymbol{x}$. We denote by $\Delta$ the Laplacian operator in  dimension $d\geq 1$ and $\sigma>0$ represents the nonlinearity exponent.  It is well known that in the \emph{sub-critical} case ($\sigma d<2$), solutions of the equation (\ref{eqNLS1}) exist and are unique globally in time for every initial condition in the Sobolev space $H^{1}(\mathbb{R}^{d})$ \cite{Weinstein83}. In contrast,  both the \emph{critical} ($\sigma d=2$) and \emph{super-critical} ($\sigma d>2$) cases admit singular (blowup) solutions, i.e., solutions that collapse in finite time (blowup time) \cite{Vlasov71}. Singular solutions are characterized by an unbounded growth of  $\nabla \psi$ close the blowup time. A negative value of Hamiltonian is sufficient for the existence of blowup \cite{Weinstein83}. A necessary condition for collapse in the critical case is that $L^2$ norm of initial conditions in $H^{1}(\mathbb{R}^{d})$ must be  equal or greater than a certain critical value $M_{\mathrm{c}}$ (depending on the dimension) \cite{Bourgain97,Weinstein83}.

During the 1980s and early 1990s, numerical simulations of singular NLS solutions with initial mass slightly above $M_c$ showed that, close the singularity, solutions can be decomposed into two components \cite{Fraiman85,Landman88,Landman91,LeMesurier88,Malkin93}. One of them, is the  collapsing core,  which approaches an  universal profile regardless of the initial condition, and the other component is the tail or outer part of the solution, which does not participate in the collapse. This fact was crucial in deriving a system of ODEs (reduced equations) in order to describe the NLS blowup dynamics \cite{Fraiman85,Landman88,LeMesurier88,Malkin93}. The first rigorous derivation was done in 2001 by Perelman for $d = 1$ and certain class of initial conditions \cite{Perelman01}, and the general case for dimensions $1 \leq d \leq 5$ was proved by Merle and Raphael  \cite{Merle03,Merle04,Merle05,Merle05_2,Merle06,Merle06_2}. 

 The physical validity of the NLS model breaks down shortly before the singularity. When NLS solutions collapse, this indicates that some of the terms neglected in the derivation become important near the singularity. In this context, different  regularization  mechanisms have been studied in order to define and understand the  solutions after the singularity time: nonlinear damping, nonlinear saturation, nonparaxiality and normal dispersion (see \cite{Fibich15} and  references therein). The existing theory to study perturbations of the NLS equation is called \emph{modulation theory}, and it was developed by Fibich and Papanicolaou \cite{Fibich98,Fibich99}. This theory is based on the adiabatic approximation. Such adiabatic approach assumes that after the singularity the collapsing core  remains close to the universal profile, by neglecting any interaction (mass transfer) between the collapsing core and the tail.

In this work we study the  NLS equation with a small damping term $\delta$ as
\begin{equation} 
 \label{eqNLS2}
    i\psi_t+ \Delta \psi + (1+i\delta)|\psi|^{4}\psi = 0, \quad \quad 0 \leq \delta \ll 1.
    \end{equation}
Equation (\ref{eqNLS2}) arises as a physical model, for example, in nonlinear optics in the setting with vanishing lower-order (quibic) nonlinear term and the nonlinear dissipation mechanism due to multi-photon absorption \cite{Fibich15,Moll03}. Similar dissipative mechanism describes four-body collisions in the Bose-Einstein condensates \cite{Ferlaino09}, and also in the context of the complex quintic Ginzburg-Landau equation \cite{Cross93}. Recently, the damped NLS (\ref{eqNLS2}) has been proposed as a mechanism for turbulent  dissipation \cite{Chung11,Josserand20}. Many numerical simulations have been carried out in order to test how well the modulation theory (adiabatic theory) describes the post-blowup dynamics in (\ref{eqNLS2}) \cite{Fibich11,Fibich12,Fibich15}. In the majority of the simulations for NLS (\ref{eqNLS2}), small perturbation of the ground state have been used as initial conditions. The present paper aims to explore numerically the post-collapse dynamics of the model (\ref{eqNLS2}), considering a generic initial condition (not too close to the ground state). Therefore, the one-dimensional equation (\ref{eqNLS2}) is solved by using the fourth-order split step method for different values of $\delta$, and with a periodic initial condition whose mass is approximately $46\%$ above the critical  one. Our findings verify the universal profile at the instant of maximum amplitude of the wavefunction. Also, the results show a qualitative agreement with the predicted exponential growth of the wave-maximum, but differing with the estimated power of $\delta$. Contrary to the prediction that a finite amount of mass is dissipated in the limit of $\delta$ going to zero for two-dimensional model \cite{Fibich01}, in one dimension, our measurements suggest that no mass is dissipated in this limit. The  breakdown of the adiabatic approximation soon after the maximum of the solution is observed. We conjecture that this invalidity could be caused by the increasing influx of mass into the collapsing core, which becomes comparable with the dissipation. In this non-adiabatic stage, a quasi-linear regime is observed. In this regime, after removing some oscillations due to the interference with the tail, the universal profile  is also verified. As a consequence, we suggest that in a free-space domain and in the limit of vanishing damping, the collapsed critical mass is instantly radiated away at the collapsing time.

The paper is organized as follows. In Section 2 we describe the model and some of its properties that are useful in the collapse dynamics.  In Section 3, we review the existing theory of blowup and post-blowup dynamics. Section 4 describes the  numerical method: the fourth-order split step method and the initial conditions. Numerical analysis  of the blowup solution and the convergence  of its inner core  to the universal profile are shown in Section 5. In that section, theoretical predictions related to the wave-maximum and wave-dissipation are analyzed as well. In the subsequent Section 6, we show numerical evidences for the breakdown of the adiabatic approximation in the post-collapse dynamics. Section 7 studies the post-adiabatic regime. We advert a quasi-linear stage after the invalidity of the adiabatic approach. We  show, however, that the universal profile in such a quasi linear regime is still valid after removing small oscillations. At the end of this section, we argue that in the limit $\delta \to 0,$ the critical mass is radiated instantly at the collapse time.  In Section 8 we provide comments and conclusions.   

\section{Model and its basic properties}
The model considered is the 1D critical damped NLS equation
 \begin{equation} 
 \label{eq1}
    i\psi_t+ \psi_{xx} + (1+i\delta)|\psi|^4\psi = 0,
    \end{equation}
where $\psi(x,t)$ is a complex-valued wave function depending on time $t \in \mathbb{R}$ and one-dimensional space variable $x \in \mathbb{R}$. In the absence of dissipation (when $\delta = 0$) NLS equation (\ref{eq1}) conserves the total mass $M$, linear momentum $P$ and Hamiltonian  $H$ defined as
\begin{equation}
M = \int|\psi|^2 dx, \quad
P = 2\int
\mathrm{Im}\left(\psi^{*}\psi_x \right)dx,\quad
H = \int\left|\psi_x\right|^2 dx -\frac{1}{3}\int|\psi|^6 dx.
\end{equation} 
 In this particular case, the minimal mass required to collapse (blowup) is  $M_{\mathrm{c}} = \sqrt{3}\pi/2$  \cite{Bourgain97,Weinstein83}.  However, the presence of nonlinear dissipation with any $\delta > 0$ regularizes the solution: it becomes unique globally in time for every initial condition in  $H^{1}(\mathbb{R})$ \cite[Theorem 33.3]{Fibich15}.

Equation (\ref{eq1}) possesses the following symmetries, which map a solution $\psi(x,t)$ into a new solution of the form 
    \begin{eqnarray}
     \textrm{space, time and phase shifts:}\quad &&  e^{i\theta}\psi(x + x_0,t+t_0),\quad
     x_0,\,t_0,\,\theta \in\mathbb{R};
    \label{eq2_sym_a}
     \\[5pt]
     \textrm{parity:}\quad &&  \psi(-x,t);
    \label{eq2_sym_b}
     \\[5pt]
     \textrm{dilation:}\quad &&
    \sqrt{\mu}\,\psi(\mu x,\mu^{2}t),\quad
     \mu > 0;
    \label{eq2_sym_c}
     \\[5pt]
     \textrm{Galilean transformation:}
     \quad &&
     \displaystyle
     \exp\left(\frac{icx}{2}-\frac{ic^2t}{4}\right)
     \psi(x - ct,t),\quad
     c \in \mathbb{R}.
    \label{eq2_sym_d}
    \end{eqnarray}
The undamped NLS (\ref{eq1}) ($\delta = 0$) admits the additional symmetries given by
    \begin{eqnarray}
     \textrm{time reversibility:}\quad&& \psi^*(x,-t);
     \label{eq2_sym2a}
     \\[5pt]
     \textrm{lens transformation:\quad}&& 
     \displaystyle
    \Psi(\xi,\tau) = \sqrt{L}\exp\left(-i\frac{L_{t}}{L}\frac{x^2}{4}\right)
     \psi(x,t),\
     \xi = \frac{x}{L(t)}, \ \tau = \int_{0}^{t}\frac{ds}{L^2(s)},
     \label{eq2_sym2b}
    \end{eqnarray}
    
\newpage
\noindent where the star denotes complex conjugation, and $L(t) = \alpha(T_c-t)$ is a linear function with an arbitrary scaling constant $\alpha > 0$ and time shift $T_c \in \mathbb{R}$.

\section{Adiabatic theory of blowup}
In the presence of weak damping, when the dissipative parameter $\delta$ is positive but small, we can distinguish three stages of the dynamics induced by the blowup phenomenon. The initial \textit{pre-blowup} dynamics follows closely the blowup behavior of system with no dissipation. In the second stage, the focusing process is stopped by dissipation with the amplitude $|\psi|$ reaching a large maximum value, and the universal amount of mass $M_{\mathrm{c}}$ concentrated in a small blowup region. Finally, the third \textit{post-blowup} stage corresponds to the reverse process of decreasing amplitude and spreading of the concentrated mass. In this section, we review the existing theoretical approach for describing this process based on the adiabatic approximation.

It is well known that, in the non-dissipative case, singular NLS solutions split into a collapsing core with the universal mass $M_{\mathrm{c}}$, and a non-collapsing tail which does not participate in the collapse \cite{Fibich99,Fibich96}. For simplicity, we consider parity-invariant (even) solutions, $\psi(x,t) = \psi(-x,t)$. Solutions with this symmetry may develop a singularity trapped at the origin, $x = 0$. Analysis of the blowup core is carried out by applying the lens transformation from (\ref{eq2_sym2b}) but with a nonlinear function $L(t)$ describing the size of  collapsing core. Substituting (\ref{eq2_sym2b}) into (\ref{eq1}), one finds that the new wave function $\Psi(\xi,\tau)$ satisfies the equation
\begin{equation}
\label{eq3_2}
    i\Psi_\tau = -\Psi_{\xi\xi} + V\Psi = 0,\quad
    V(\xi,\tau) = -\frac{\beta\xi^2}{4}- (1+i\delta)|\Psi|^4,
\end{equation}
where the nonlinear function $L(t)$ leads to an extra quadratic potential term, $\beta\xi^2/4$, with
    \begin{equation}
    \beta = -L^{3}L_{tt}.
    \label{eq3_3}
    \end{equation}
Equation (\ref{eq3_2}) describes the dynamics in new (rescaled) spatial and temporal variables, $\xi$ and $\tau$. When $\delta = 0$, the blowup dynamics in new variables corresponds to the vanishing positive $\beta \searrow 0$ as $\tau \to +\infty$. When $\beta \equiv 0,$ the renormalized equation (\ref{eq3_2}) admits a solution of the form 
    \begin{equation}
    \label{eq3_4}
    \Psi(\xi,\tau) = e^{-i\tau}R(\xi),
    \end{equation}
where $R(\xi)$ solves the boundary value problem
    \begin{equation}
    \label{eq3_5}
    R_{\xi\xi}-R+R^5 = 0, \quad 
    R_\xi(0)=0, \quad \lim_{\xi \to +\infty}R(\xi) = 0.
    \end{equation}
It is found explicitly as
\begin{equation}
\label{GroundState}
    R(\xi) = \frac{3^{1/4}}{\cosh^{1/2}(2\xi)}.
\end{equation}
When $\beta$ is small and positive, the form of potential $V$ from (\ref{eq3_2}) leads to the tunneling effect. This effect is described in the leading order by the equation of the form \cite{Landman88,LeMesurier88,Malkin93,Sulem99,Dyachenko92}
    \begin{equation}
    \label{eq3_6}
    \beta_\tau = 
    -\nu(\beta),\quad \nu(\beta):=\left\{\begin{array}{ll}
         c_{\beta}e^{-\pi/\sqrt{\beta}},& \beta > 0,\\[3pt]
            0,& \beta < 0, 
            \end{array}\right.
    \end{equation}
where $c_\beta = 1024/\pi^{3}$ is the universal coefficient.
This derivation is based on the adiabatic theorem, which is applicable due to a slow change of $\beta(\tau)$, and on the fact that the ground state with energy $E = -1$ becomes a resonant (complex energy) state, whose small imaginary part describes a decay due to the tunneling effect \cite{Dyachenko92}.

Returning now to the dissipative equation (\ref{eq3_2}) with small positive $\delta$, the dissipative effect can be taken into account using the perturbation theory. In the framework of the adiabatic approximation, it yields a leading order correction to the equation for $\beta$ as \cite{Fibich98,Fibich99,Passot05}
    \begin{equation}
    \beta_\tau = -\nu(\beta)-c_d\delta, 
    \label{eq3_8}
    \end{equation}
with the coefficient $c_d = 192/\pi^2$.
Using relations (\ref{eq2_sym2b}), one can write the resulting adiabatic approximation (universal profile) for the blowup solution in the dissipative system as
    \begin{equation}
    \psi(x,t) = \frac{R(\xi)}{\sqrt{L(t)}}
    \exp\left(i\tau(t)+i\frac{L_t}{L}\frac{x^2}{4}\right),  \label{eq3_10}
    \end{equation}
where the functions $L(t)$, $\tau(t)$ along with $\beta(t)$ are determined by the reduced system of three ordinary differential equations
    \begin{equation}    
    \label{eq3_11}
    L_{tt} = -L^{-3}\beta(t),\quad
    \tau_t = L^{-2},\quad
    \beta_t = -L^{-2}\nu(\beta)
    -c_dL^{-2}\delta,
    \end{equation}
as can be seen from the equations (\ref{eq2_sym2b}), (\ref{eq3_3}) and (\ref{eq3_8}).

\section{Numerical method}

Our analysis is based on high-accuracy numerical simulations of equation (\ref{eq1}) in the periodic domain $x \in [-\pi,\pi],$ and produced by applying the fourth-order split step method similar to the one described in \cite{Chung11,Yoshida90}. 
For this purpose, equation (\ref{eq1}) is written as
\begin{equation}
\label{Num1}
\psi_{t} = (\hat{L}+\hat{N})\psi,
\end{equation}
where $\hat{L}$ and $\hat{N}$ are the linear and nonlinear operators defined as
    \begin{equation}
    \hat{L}\psi = i\psi_{xx}, \quad \quad \quad \hat{N}\psi = i(1+i\delta)|\psi|^{4}\psi. 
    \end{equation}
Given the time step $\Delta t$, the solution $\psi(t + \Delta t) = e^{\Delta t(\hat{N}+\hat{L})}\psi(t)$ is approximated by the expression \cite{Chung11,Yoshida90}
    \begin{equation}
    \psi(t + \Delta t) = e^{c_{1}\Delta t \hat{N}}e^{d_{1}\Delta t \hat{L}} e^{c_{2}\Delta t \hat{N}}e^{d_{2}\Delta t \hat{L}} e^{c_{2}\Delta t \hat{N}}e^{d_{1}\Delta t \hat{L}}e^{c_{1}\Delta t \hat{N}}  \psi(t),
    \end{equation}
where
    \begin{equation}
    c_{1} = \frac{1}{2(2-2^{1/3})}, \quad   c_{2} = \frac{1-2^{1/3}}{2(2-2^{1/3})}, \quad 
    d_{1} = \frac{1}{2-2^{1/3}}, \quad d_{2} = \frac{-2^{1/3}}{2-2^{1/3}}.
    \end{equation}
The nonlinear part $e^{c_{j}\Delta t \hat{N}}$ is approximated by
\begin{equation}
    e^{c_{j}\Delta t \hat{N}}  \psi(t) \approx e^{ic_{j}\Delta t (1+i\delta)|\psi(t)|^{4} }  \psi(t),  \quad j = 1,2.
\end{equation}
For the linear part $e^{c_{j}\Delta t \hat{L}}$, the  exact expression
   \begin{equation}
       e^{c_{j}\Delta t \hat{L}}  \psi(t) = \mathcal{F}^{-1}\left[
       e^{-ic_{j}\Delta t k^2}\mathcal{F}[ \psi(t)]\right],   \quad j = 1,2,
   \end{equation}
was computed by using the fast Fourier transform (FFT) algorithm implemented in MATLAB.

We started the simulations with the spatial grid size $\Delta x = 2\pi/2^{12}$. It was decreased by two with the Fourier interpolation for the solution each time, when the spectrum was reaching the largest wavenumbers, so that the discretization error was kept at the level of round-off noise. At the end of simulations, we reached the resolution up to $2^{23}$ points. For the time step, we used the relation $\Delta t = 0.2(\Delta x)^{2}/\pi$ in order to avoid numerical instability \cite{Chung11}.

In the numerical simulations we used the initial condition $\psi_{0}(x)= 0.6\left[1+\cos^{8}(x/2) +0.1i\right]$ and the values of $\delta = 10^{-2},5\times 10^{-3},2.5\times 10^{-3},10^{-3},5\times 10^{-4}$. Such initial condition has initial mass $M(0) \approx 3.96577$ which corresponds to $ 46\%$ above the critical mass. For the undamped  solution ($\delta = 0$) one has $H(0) \approx -0.6888$. Therefore, it blows up, and the critical time is estimated numerically as $T_{c} \approx 1.4826$. The linear momentum is $P(0) = 0.$

\section{Blowup dynamics}

In this section, we will verify the regularization mechanism in the damped NLS, and analyze that the core of the solution  approaches  the universal profile in the collapse dynamics. Also, theoretical predictions related to the wave-maximum and wave-dissipation are considered. Thus, this section mainly verifies the existing theory and provides some new observations regarding the total dissipation at the collapse.

\begin{figure}[H]
	\centering
	\subfloat[]{\includegraphics[width=0.33\textwidth]{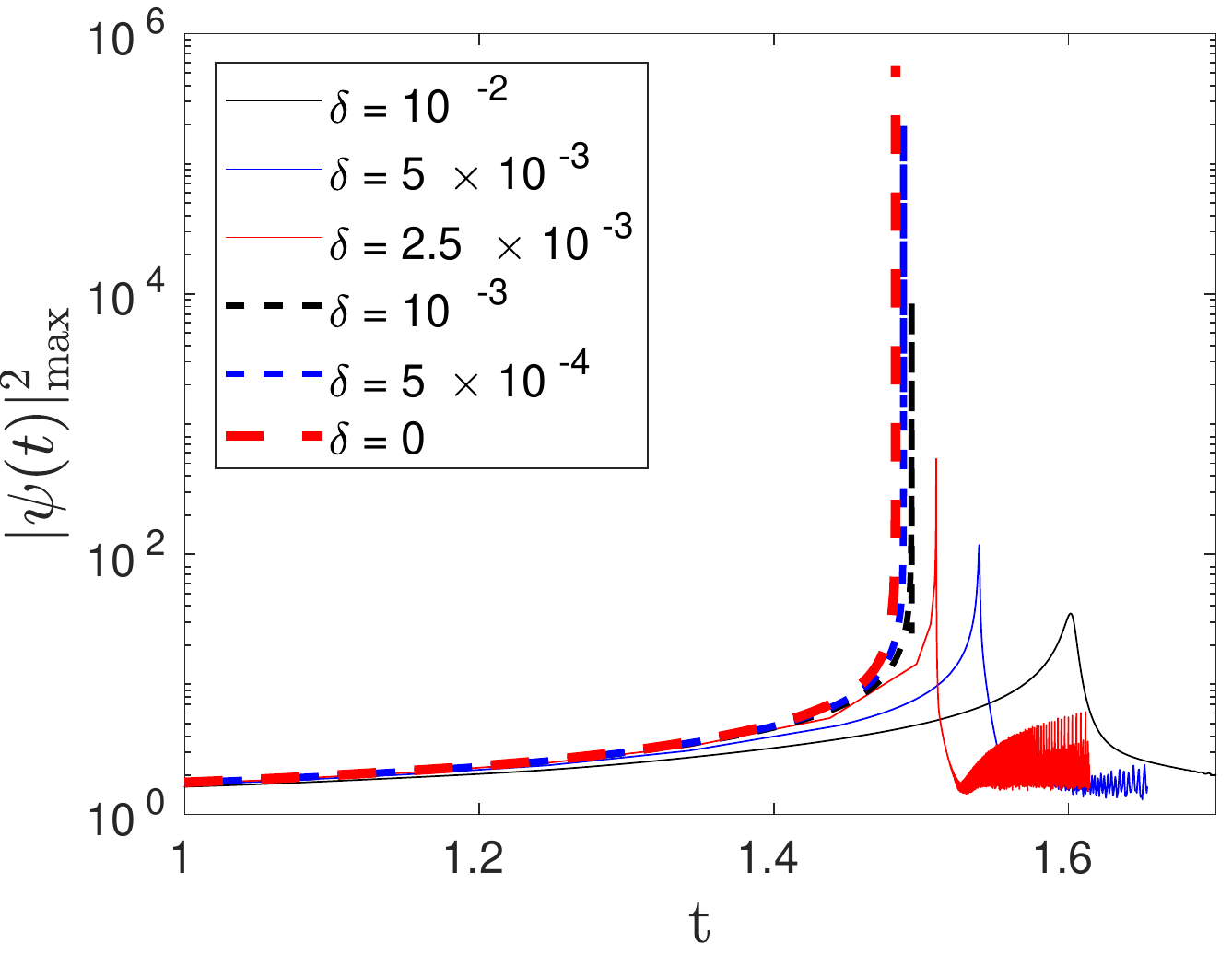}}
	\hspace{0.2cm}
	\subfloat[]{\includegraphics[width=0.33\textwidth]{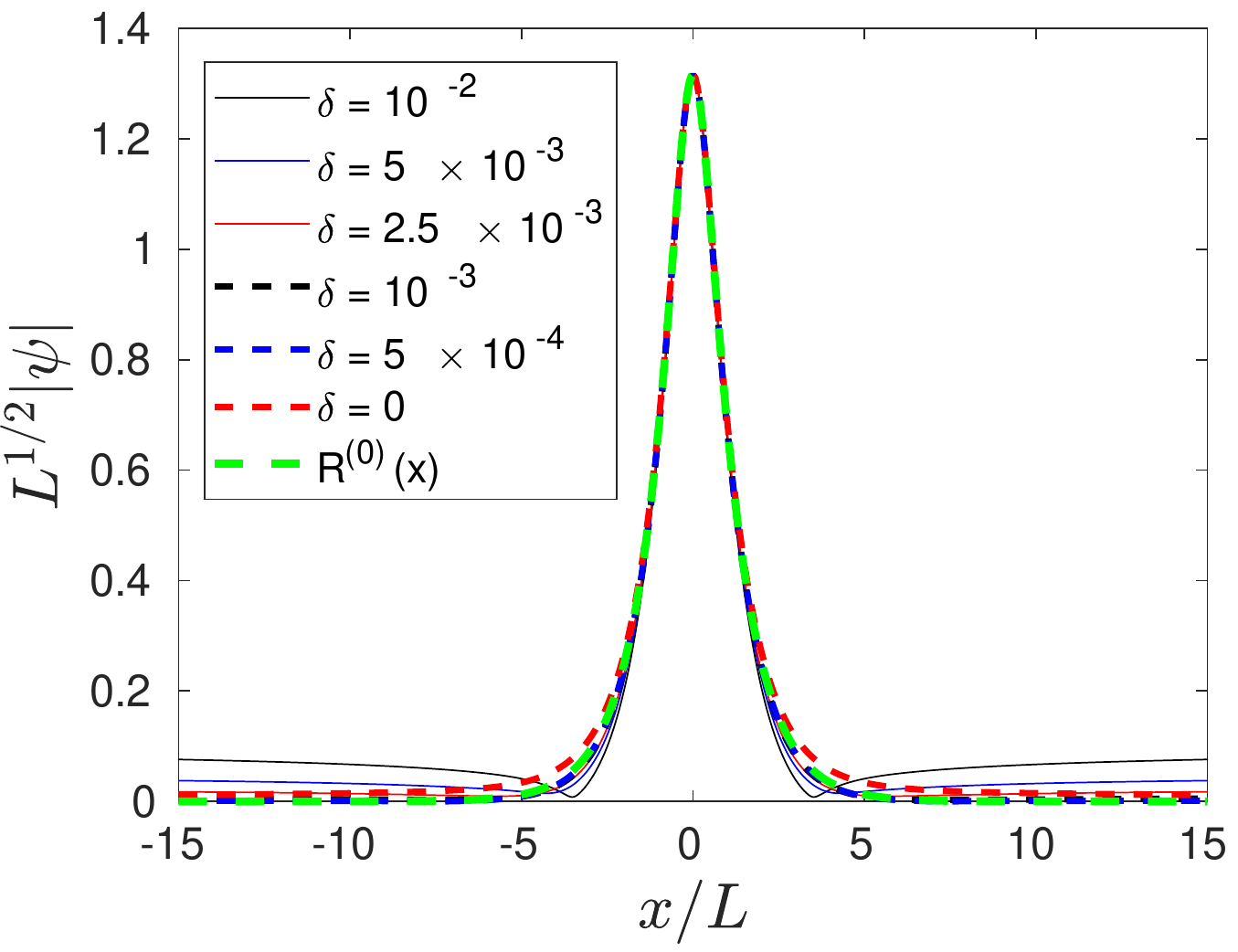}}
	\hspace{0.2cm}
	\subfloat[]{\includegraphics[width=0.3\textwidth]{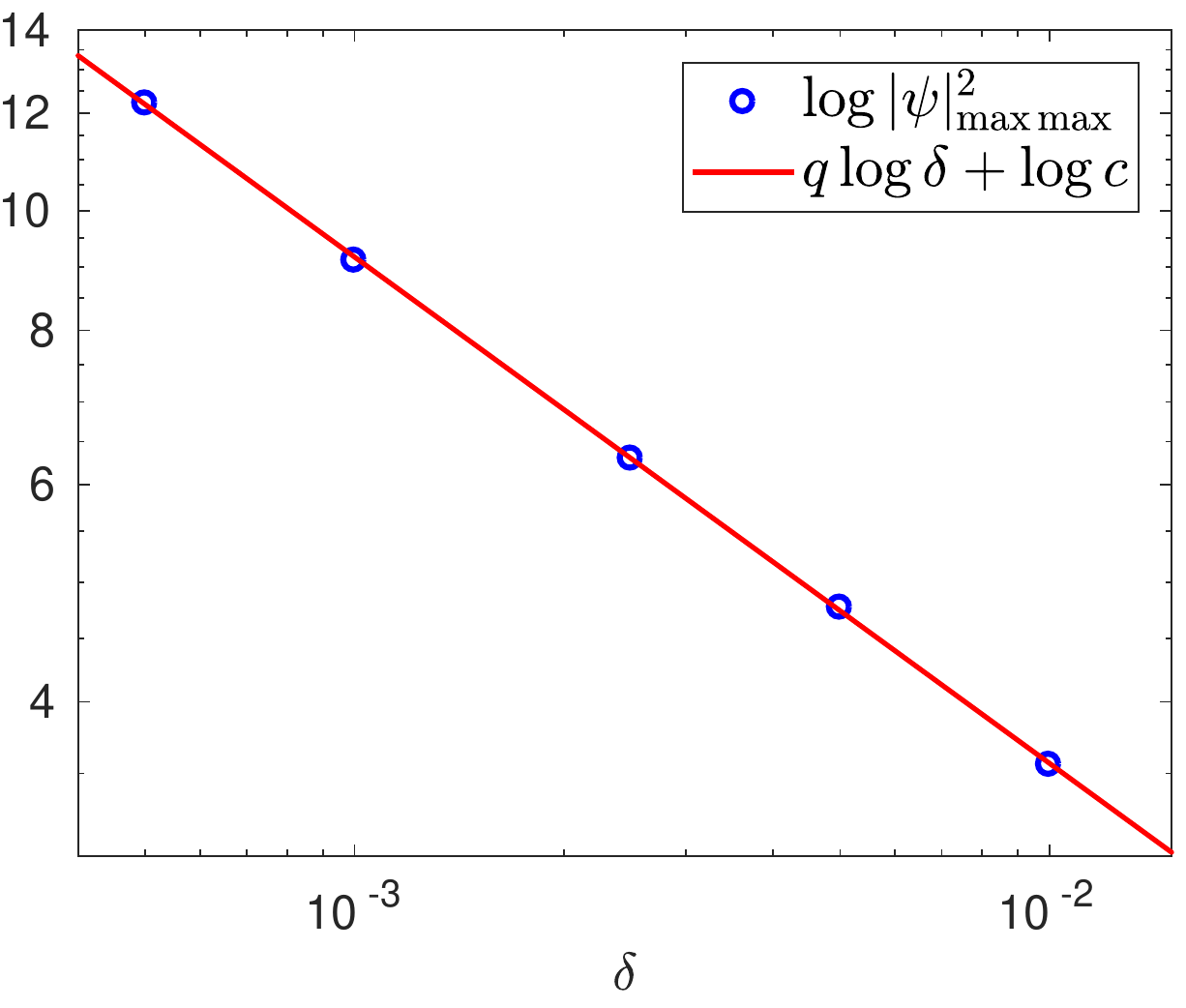}}
	\hspace{0.2cm}
	\subfloat[]{\includegraphics[width=0.3\textwidth]{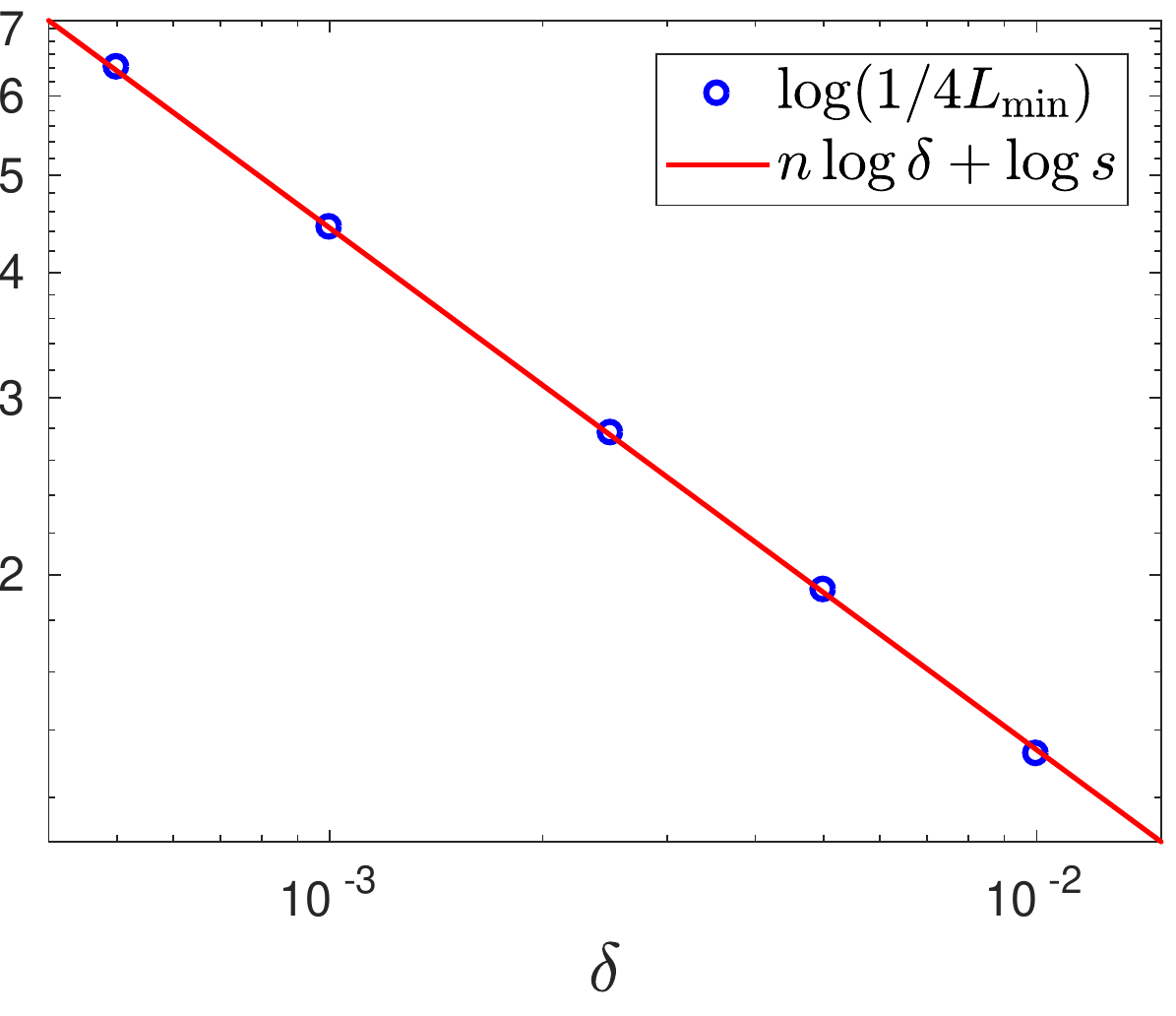}}
 \subfloat[]{\includegraphics[width=0.33\textwidth]{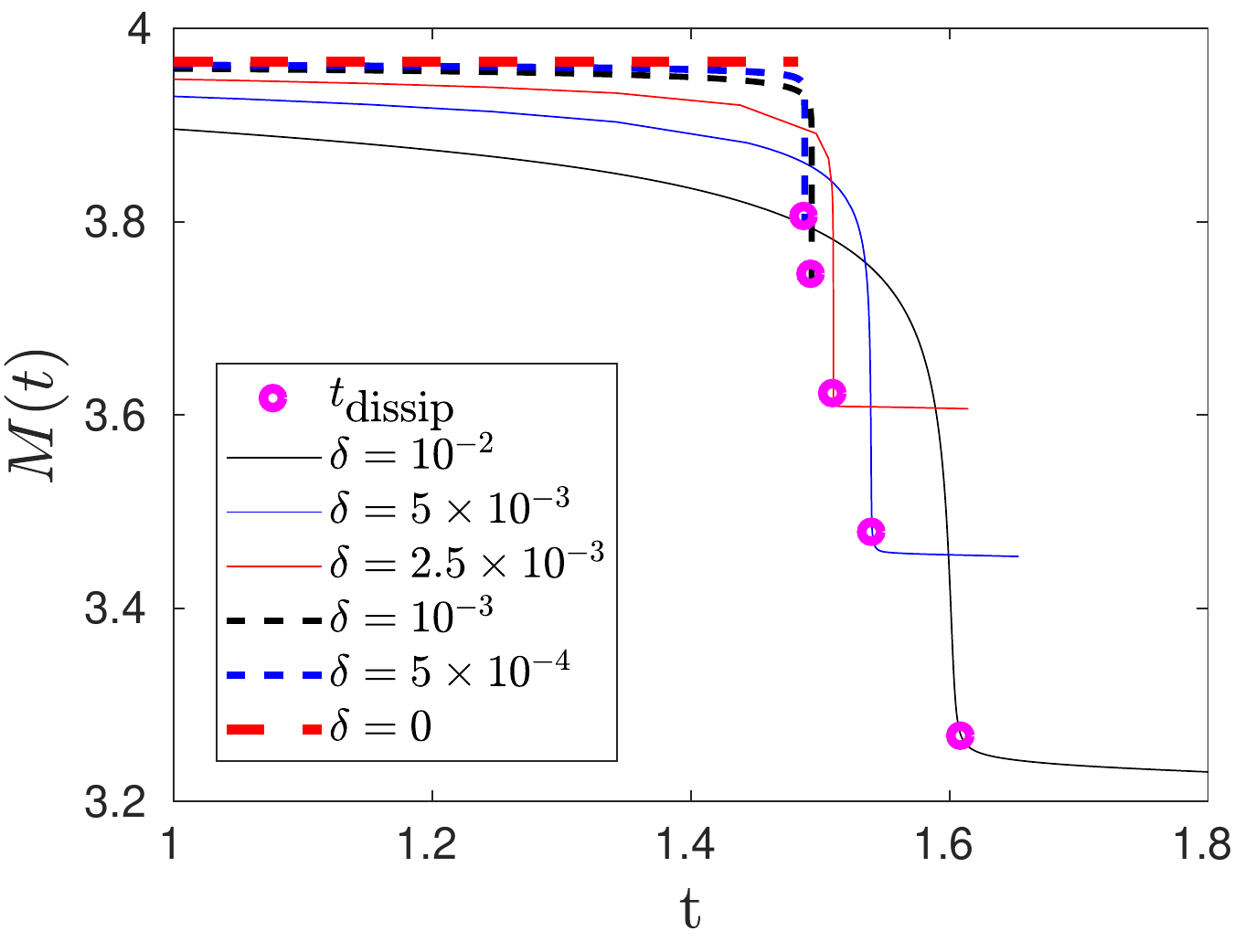}}
	\hspace{0.2cm}
 \subfloat[]{\includegraphics[width=0.3\textwidth]{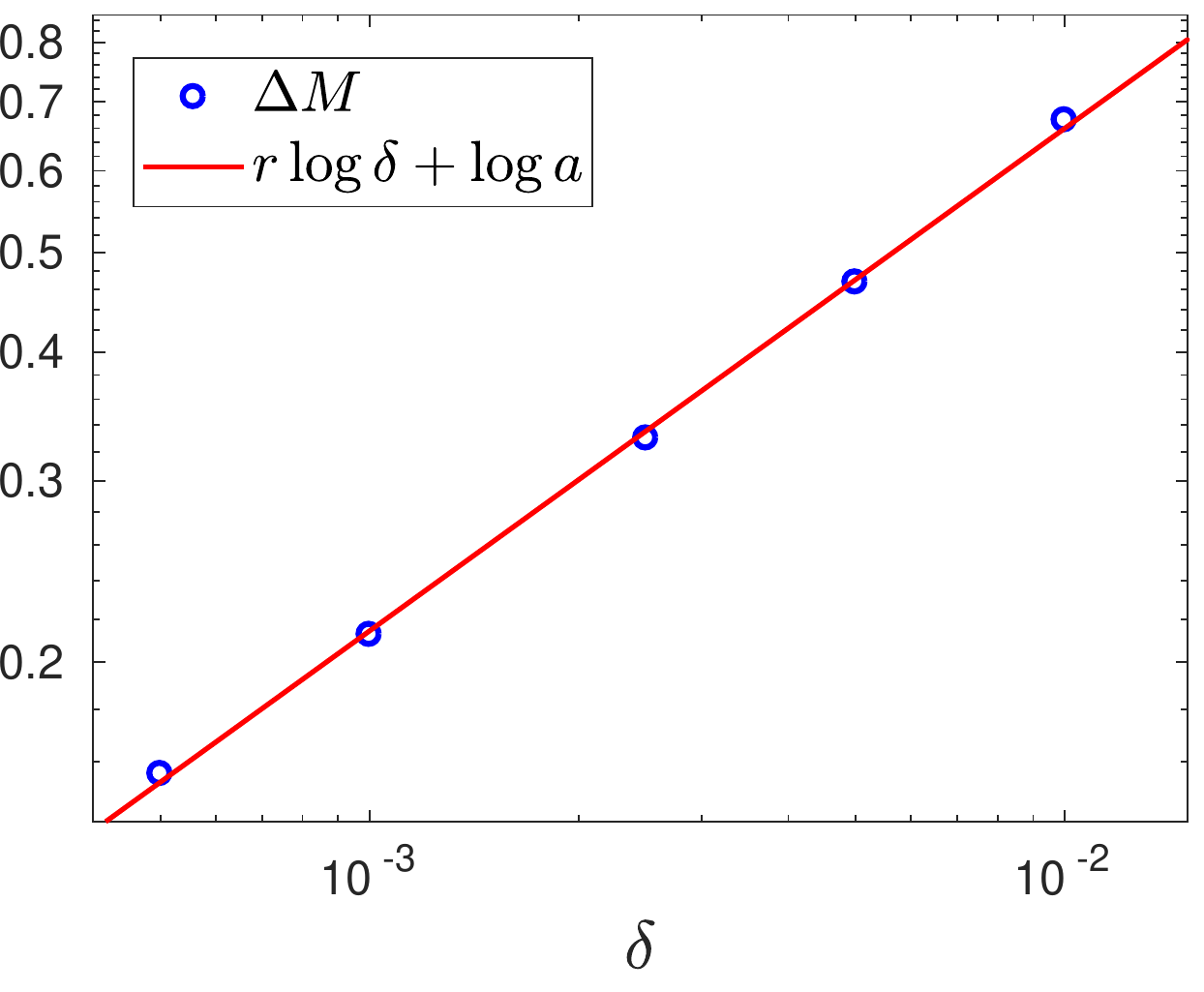}}
	\hspace{0.2cm}
	\caption{(a) Evolution of $|\psi(t)|_{\max}^{2}$ for different values of $\delta$. It shows that the dissipation term regularizes the solution. (b) Comparison between the rescaled profile $L^{1/2}|\psi(x/L,t)|$ of the dissipative solution and the ground state $R^{(0)}(x)$ at the time $t_{\max}$. (c) $\log |\psi|_{\max \max}^2$  as function of $\delta$ in a log-log scale. It shows that  $|\psi|_{\max \max}^2 \sim \exp(c\delta^{q})$ with $c\approx 0.54$ and $q \approx -0.41$.  (d) Solution of the reduced system (\ref{eq3_11}) for $\delta = 10^{-2},5\times 10^{-3}, 2.5\times 10^{-3}, 10^{-3}, 5\times 10^{-4}$ and initial conditions corresponding to the blowup solution at time $t \approx 1.4$, i.e., $L(1.4) \approx 0.3112,$ $\beta(1.4) \approx 0.4473$ and $L_{t}(1.4) \approx -2.6454.$  The maximum of  $1/L(t)$ satisfies the law $1/L_{\min} \sim 4\exp(s \delta^{n})$ with $s \approx 0.122$ and $n \approx -0.52$. (e) Evolution of mass $M(t)$ for each $\delta.$ The time $t_{\textrm{dissip}}$ corresponds to the moment where the  derivative of $M(t)$ is approximately $10\%$ of its maximum absolute value. (f) Amount of mass dissipated within a collapse $\Delta M = M(0)-M(t_{\textrm{dissip}})$. Numerical simulations support the scaling $\Delta M  \sim a \delta^{r}$ with $a \approx 6.59$ and $r \approx 0.49.$ Consequently, our result indicate that in the limit $\delta \to 0^+$ no mass is dissipated in the collapse.}
\label{Fig1}
\end{figure}

Figure \ref{Fig1}(a) plots the maxima of the squared modulus of the wavefunction $\psi$ depending on time for different damping coefficients $\delta$, i.e., $|\psi(t)|_{\max}^{2} = \max_{x}|\psi(x,t)|^2$. One can observe, as it is expected, that after adding the small damping term to the NLS equation, the singularity is cured and  the solution continues  after the collapse. A  convergence to the unperturbed NLS, $\delta = 0$, is  showed in the same figure.  Notice that the peak is located earlier and its amplitude is larger as $\delta$ decreases. 

In order to verify the ansatz (\ref{eq3_10}) within the collapse dynamics, we  compared the rescaled profile $L^{1/2}|\psi(x/L,t)|$ with the ground state $R^{(0)}(x)$ at the instant of maximum amplitude of the solution $t_{\max}$ (and $T_c$ for $\delta = 0$), as can be seen in Figure \ref{Fig1}(b). Using equation (\ref{eq3_10}) at $x = 0,$ we  computed the width of the solution, $L(t)$, by the relation
\begin{eqnarray}
\label{eq_L}
 L(t) = \frac{\sqrt{3}}{|\psi(0,t)|^2}.
\end{eqnarray}
As displayed in Figure \ref{Fig1}(b), the quasi self-similar collapsing core  becomes closer of $R^{(0)}(x)$ for smaller $\delta>0$. One can see in the same figure, that the collapsing core corresponding to the damped coefficient $\delta = 5\times 10^{-4}$ is closer to the ground state $R^{(0)}(x)$ than the undamped case $\delta = 0.$ This behaviour can be justified by the physical meaning of $\beta(t)$, which  is proportional to the difference between the mass of the inner core of the solution and the critical mass $M_c$ \cite{Malkin93,Fibich15}. In our  simulations $|\beta| \approx 0.0943$ and $|\beta| \approx 0.0470$ for the damping coefficients $\delta = 0$ and $\delta = 5\times 10^{-4}$ respectively, therefore, we expect that the collapsing core of damped solution with $\delta = 5\times 10^{-4}$ should be closer to the ground state than undamped solution with $\delta = 0.$

Now, in Figures \ref{Fig1}(c-d) we analyze the $\delta$-dependence of the maximum of $|\psi|^2$, i.e., $|\psi|_{\max \max}^{2} = \max_{x,t}|\psi(x,t)|^2.$ The first of these figures describes the dynamics of $|\psi|_{\max \max}^{2}$, and indicates the asymptotic law 
\begin{eqnarray}
\label{PowerLaw1}
 |\psi|_{\max \max}^2 \sim \exp(c\delta^{q}) \quad \textrm{where} \quad  c\approx 0.54, \quad  \quad q \approx -0.41. 
\end{eqnarray}
This observation can be compared with the theoretical prediction given by the reduced equations (\ref{eq3_11}). For that purpose, reduced equations (\ref{eq3_11}) have been solved numerically for various  $\delta$ using  

\noindent the initial conditions $L(1.4) \approx 0.3112,$ $\beta(1.4) \approx 0.4473$ and $L_{t}(1.4) \approx -2.6454,$ obtained from the undamped NLS. The result is displayed in the Figure \ref{Fig1}(d), and it suggests that 
the maximum of the function $1/L(t)$, $1/L_{\min},$ behaves following the law 
\begin{eqnarray}
 1/L_{\min} \sim 4\exp(s \delta^{n}) \quad \textrm{with} \quad s \approx 0.12,  \quad  \quad n \approx -0.52.
\end{eqnarray}
Consequently, by the relation (\ref{eq_L}) we get
\begin{eqnarray}
\label{PowerLaw2}
 |\psi|_{\max \max}^2 \sim 4\sqrt{3}\exp(s\delta^{n}) \quad \textrm{where} \quad  s\approx 0.122, \quad  \quad n \approx -0.52. 
\end{eqnarray}
The  exponential growth  established in both  (\ref{PowerLaw1}) and (\ref{PowerLaw2}) have qualitative agreement with the theoretical prediction, see, e.g., \cite{Fibich11,Fibich98_2}. However, the estimates for the power of $\delta$ in (\ref{PowerLaw1}) and (\ref{PowerLaw2}) are different from each other and from the theoretical estimate $-1$ \cite{Fibich11,Fibich98_2}. This discrepancy questions the validity of reduced equations at times around the maximum of the amplitude.

Finally, in this section we analyze the wave-dissipation dynamics. In Figure \ref{Fig1}(e) we  plotted the time evolution of the mass $M(t)$ for various $\delta$, and  we can observe, as it is expected, that dissipation becomes important only in the collapse event. In these terms, the amount of dissipated mass is defined as 
\begin{eqnarray}
 \Delta M = M(0)-M(t_{\textrm{dissip}}),
\end{eqnarray}
where  $t_{\textrm{dissip}}$ is a time after $t_{\max}$ (see Figures \ref{Fig4}(a-c)), and corresponding in our study to the instant where the dissipative effect is almost unimportant. It was defined as the time where  the derivative of $M(t)$ reaches approximately the  $10\%$  of its maximum absolute value; see Figure \ref{Fig1}(e). The result of our analysis is presented in the Figure \ref{Fig1}(f) in a log-log scale, and it shows that $\Delta M$  scales like
\begin{eqnarray}
\label{823}
 \Delta M \sim a\delta^r \quad \textrm{with} \quad a \approx 6.59, \quad \quad  r \approx 0.49.
\end{eqnarray}
We point out that, taking different percentages in the definition of $t_{\textrm{dissip}}$ do not change the power $r$ of the scaling (\ref{823}).  Therefore, the power law (\ref{823}) indicates that in the limit of vanishing damping no mass is dissipated in the collapse. This fact is opposite to the theoretical prediction established  for the two-dimensional damped NLS \cite{Fibich01}, in which the limiting mass loss is non-zero.

\section{Breakdown of adiabatic approximation}

Adiabatic approximation is based on the fact that  the shape of collapsing core of the solution remains close to the ground state (\ref{GroundState}), and the potential $V(\xi,\tau)$ in equation (\ref{eq3_2}) varies slowly in terms of the rescaled time $\tau$. In these terms, one expects that adiabatic approach can be also used for describing post-blowup dynamics. 

\begin{figure}[H]
	\centering
	\subfloat[]{\includegraphics[width=0.32\textwidth]{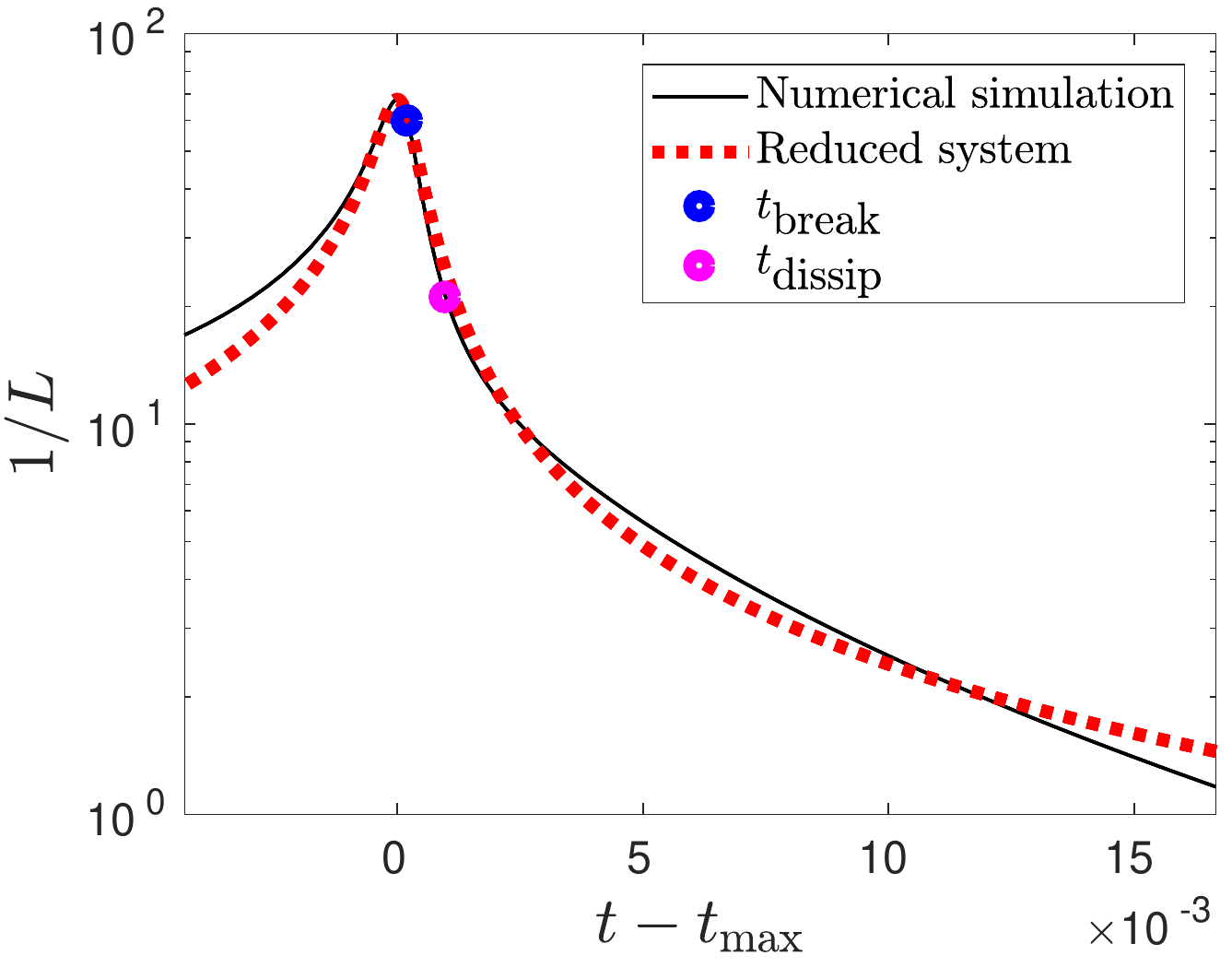}}
	\hspace{0.2cm}
	\subfloat[]{\includegraphics[width=0.32\textwidth]{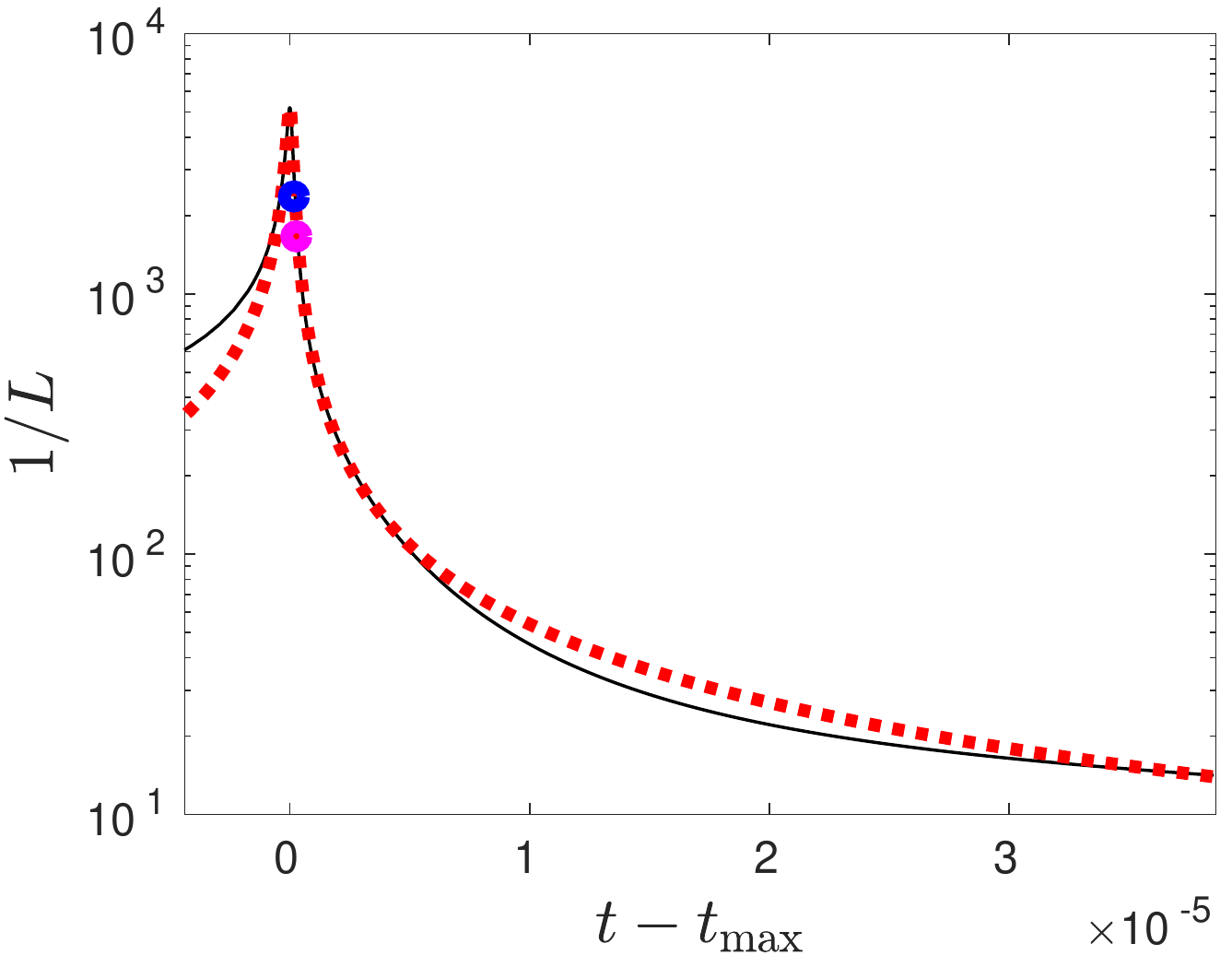}}
	\hspace{0.2cm}
	\subfloat[]{\includegraphics[width=0.32\textwidth]{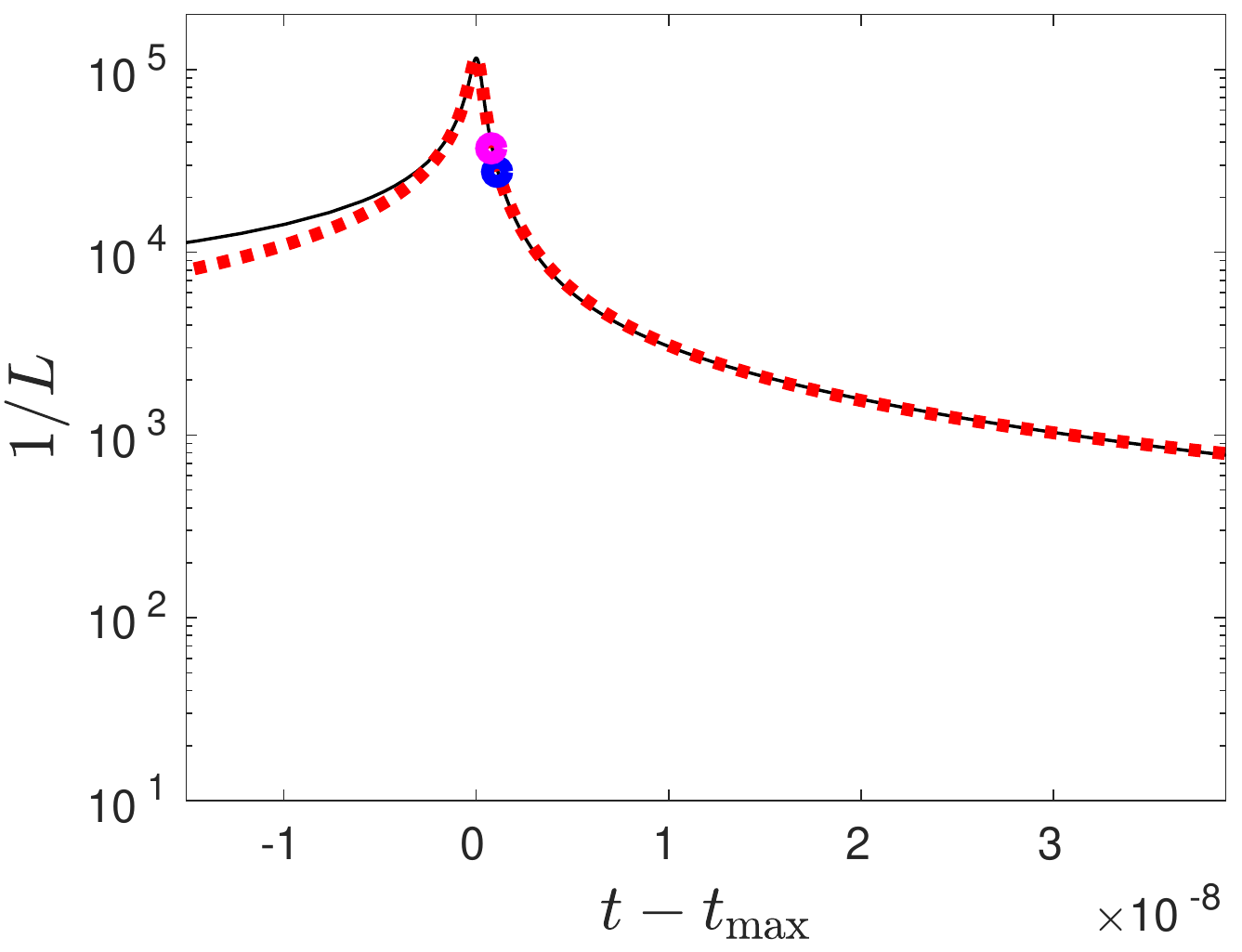}}
	\hspace{0.2cm}
	\subfloat[]{\includegraphics[width=0.32\textwidth]{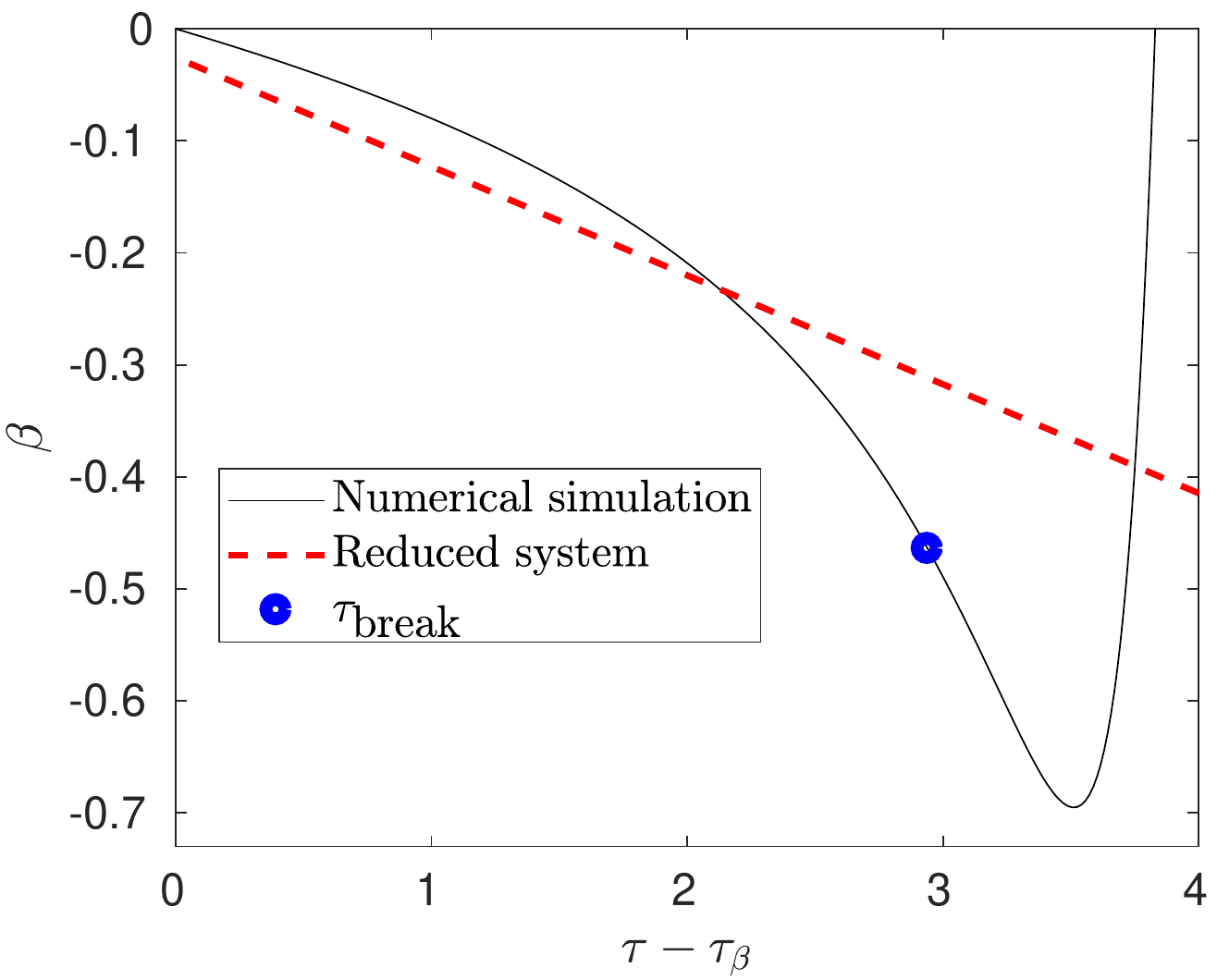}}
 \subfloat[]{\includegraphics[width=0.33\textwidth]{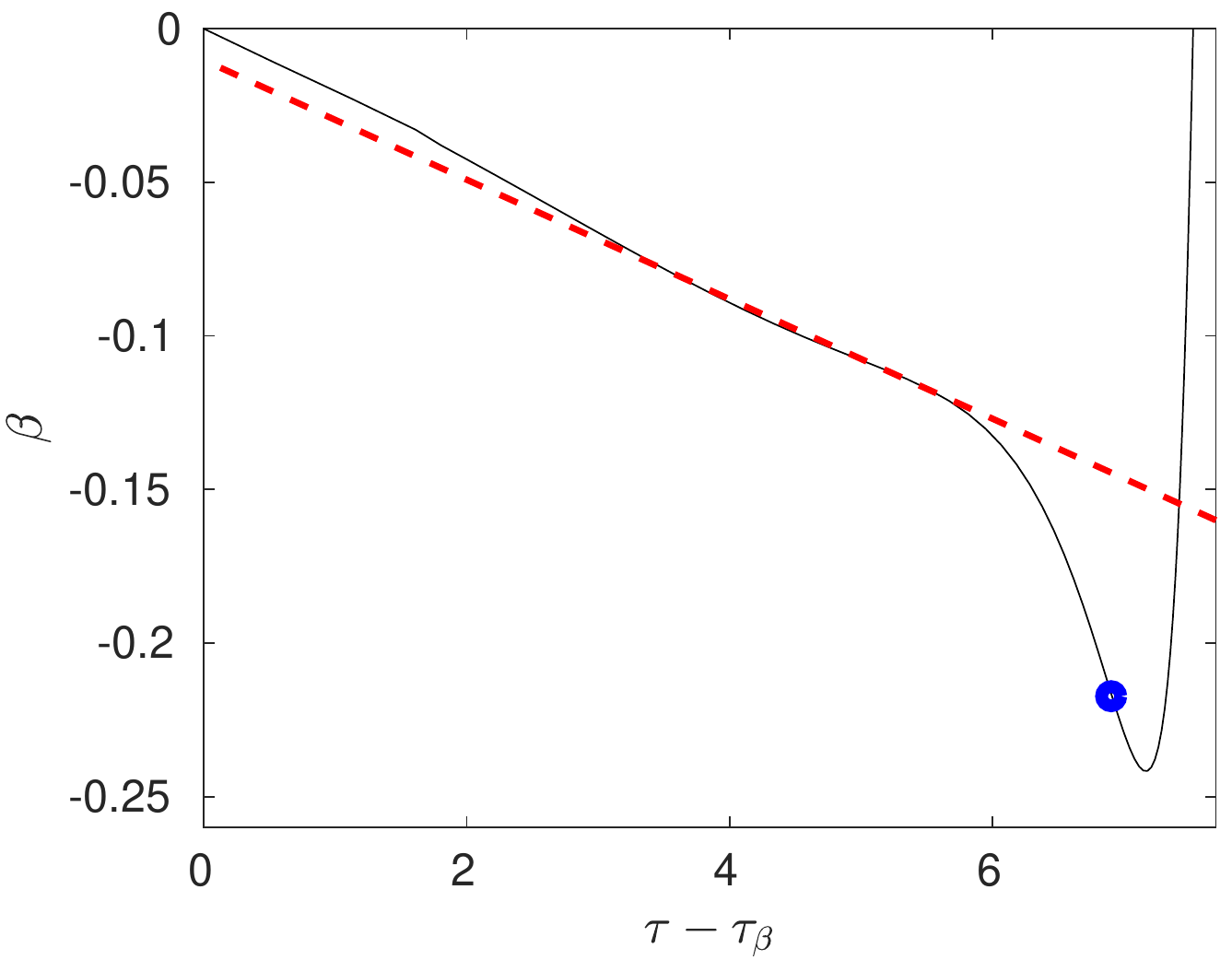}}
	\hspace{0.2cm}
 \subfloat[]{\includegraphics[width=0.33\textwidth]{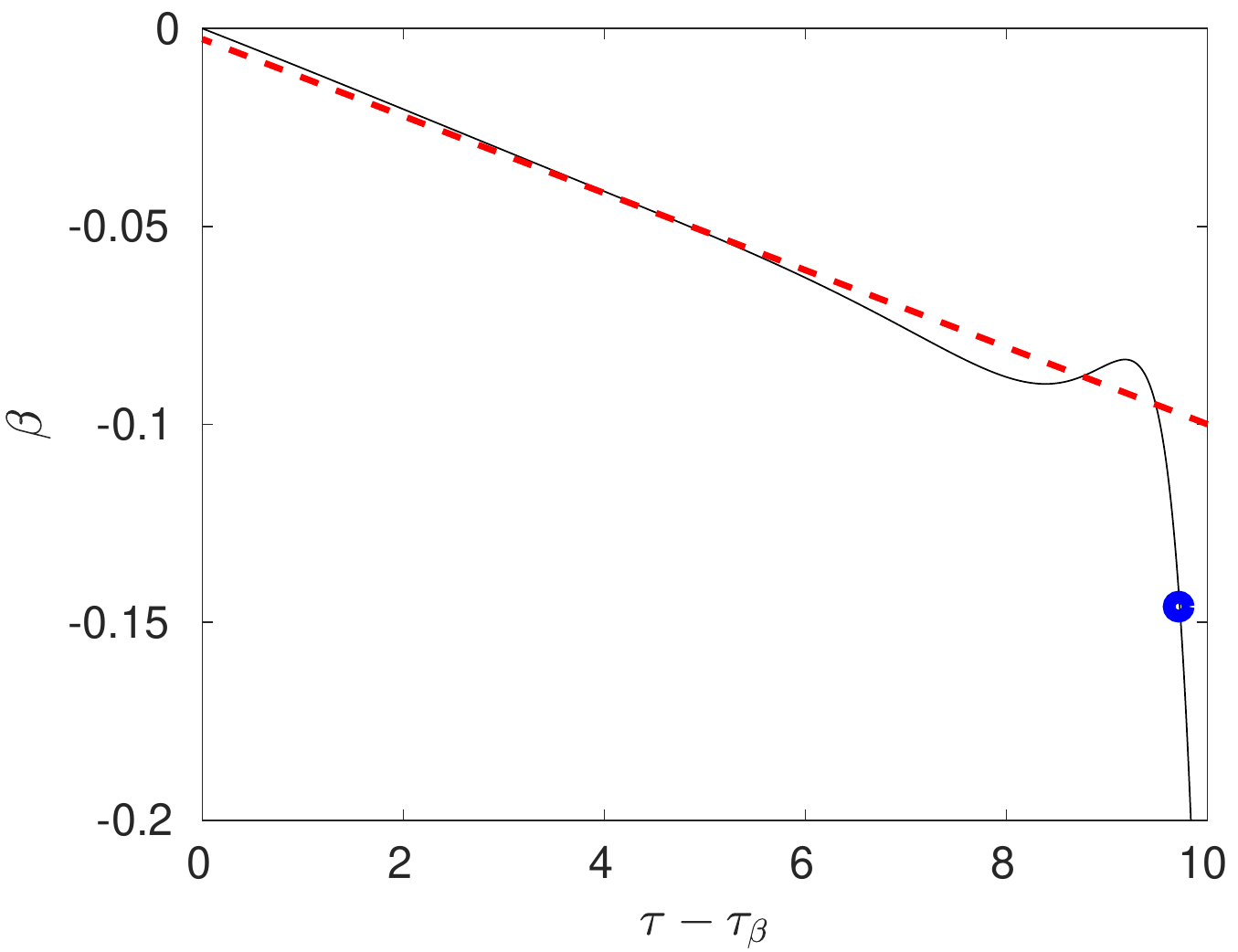}}
	\hspace{0.2cm}
	\caption{Comparison between the solution of the reduced system (\ref{eq3_11})  and  direct numerical simulations of the damped NLS for $\delta = 5 \times 10^{-3}$,  $\delta = 10^{-3}$ and $\delta = 5 \times 10^{-4}$. In Figures \ref{Fig4}(a-c), one sees a reasonable agreements among the functions $1/L(t)$ in a neighborhood of $t_{\max}$ for $\delta$ decreasing. In contrast, in Figures \ref{Fig4}(d-f) we have plotted the functions $\beta(\tau)$. In such figures, one observes that $\beta(\tau)$, as $\delta$ goes down, becomes closer to the linear behavior predicted by the adiabatic approximation at early times. But, after a certain  time, a prominent  deviation from the linear approximation is noticed, which implies the breakdown of the adiabatic approach. Therefore, we defined the break down time $\tau_{\textrm{break}}$, as the time where the difference between the function $\beta(\tau)$ and linear approximation is around $50\%$. The corresponding location of $t_{\textrm{break}}$ on the function $1/L(t)$ in panels (a-c) adverts the invalidity of the adiabatic approach shortly after the peak.}
\label{Fig4}
\end{figure}
\noindent Fibich and Klein \cite{Fibich11} considered the explicit blowup solution, and by using the reduced equations  (\ref{eq3_11}), calculated the nonlinear damping continuation in the limit $\delta \to 0$. The case of solutions with mass above $M_c$  was also addressed  in \cite{Fibich11}, taking as initial condition a small perturbation of the explicit blowup solution whose mass was approximately $2.5\%$ above the critical one. Numerical simulations carried out in \cite{Fibich11} showed that the first moments in the post-blowup dynamics of the damped NLS equation (\ref{eq1}) are described by the adiabatic approximation as long as $\delta$ is small. Consequently, such simulations showed a clear evidence of the breakdown of the adiabatic stage  shortly after the arrest of collapse.  In the present section, our intention is to see how the adiabatic approximation works for a generic initial condition, not too close to the ground state (\ref{GroundState}). Therefore, we will compare the solution of the reduced system (\ref{eq3_11}) (functions $1/L$ and $\beta$) with  our numerical simulations of the NLS (\ref{eq1}). We arrive to the same conclusion on the breakdown of the adiabatic approximation, and propose a plausible explanation.

 The direct  simulation of the NLS (\ref{eq1}), provide us the functions $L(t)$ and $\beta(t)$  through the relations  (\ref{eq_L}) and (\ref{eq3_3}) respectively. The variable $\tau(t)$ is obtained by $\tau = \arg \psi(0,t)$ as follows from equation (\ref{eq3_10}). The minimum value of the function $L(t)$ computed previously, and corresponding $\beta$ and $\tau$, were taken as initial condition.  For this initial point, we solved the reduced system forwards and backwards. The results are presented in Figure \ref{Fig4}, which plots the functions $1/L(t)$ and $\beta(\tau)$  for various $\delta$. Figures \ref{Fig4}(a-c) show a quantitative agreement of the function $1/L(t)$ in a vicinity of $t_{\max}$ improving for smaller damping coefficient. Figures \ref{Fig4}(d-f) display  $\beta(\tau)$ as a function of $\tau - \tau_{\beta}$ where $\tau_{\beta}$ is chosen such that $\beta(\tau_{\beta}) = 0$. We see that $\beta(\tau)$ approaches the linear behaviour, but after a certain moment, starts to deviate from the linear dynamics forecasted by the adiabatic approximation. This deviation manifests the breakdown of the adiabatic approximation.

We  define the breakdown time $\tau_{\textrm{break}}$, and consequently $t_{\textrm{break}}$,  as the time where the difference between the function $\beta(\tau)$ and linear approximation exceed $50\%$. The times $t_{\textrm{break}}$ and $\tau_{\textrm{break}}$ are plotted by blue dots in the Figure \ref{Fig4}. Therefore, our numerical simulations indicate that breakdown of the adiabatic approach occurs  shortly after the peak, and the time-interval in which it is valid collapses with $\delta$ going to zero, see Figures \ref{Fig4}(a-c). It is in concordance with what was reported in \cite{Fibich11,Fibich12}.

\begin{figure}[H]
	\centering
	\subfloat[]{\includegraphics[width=0.32\textwidth]{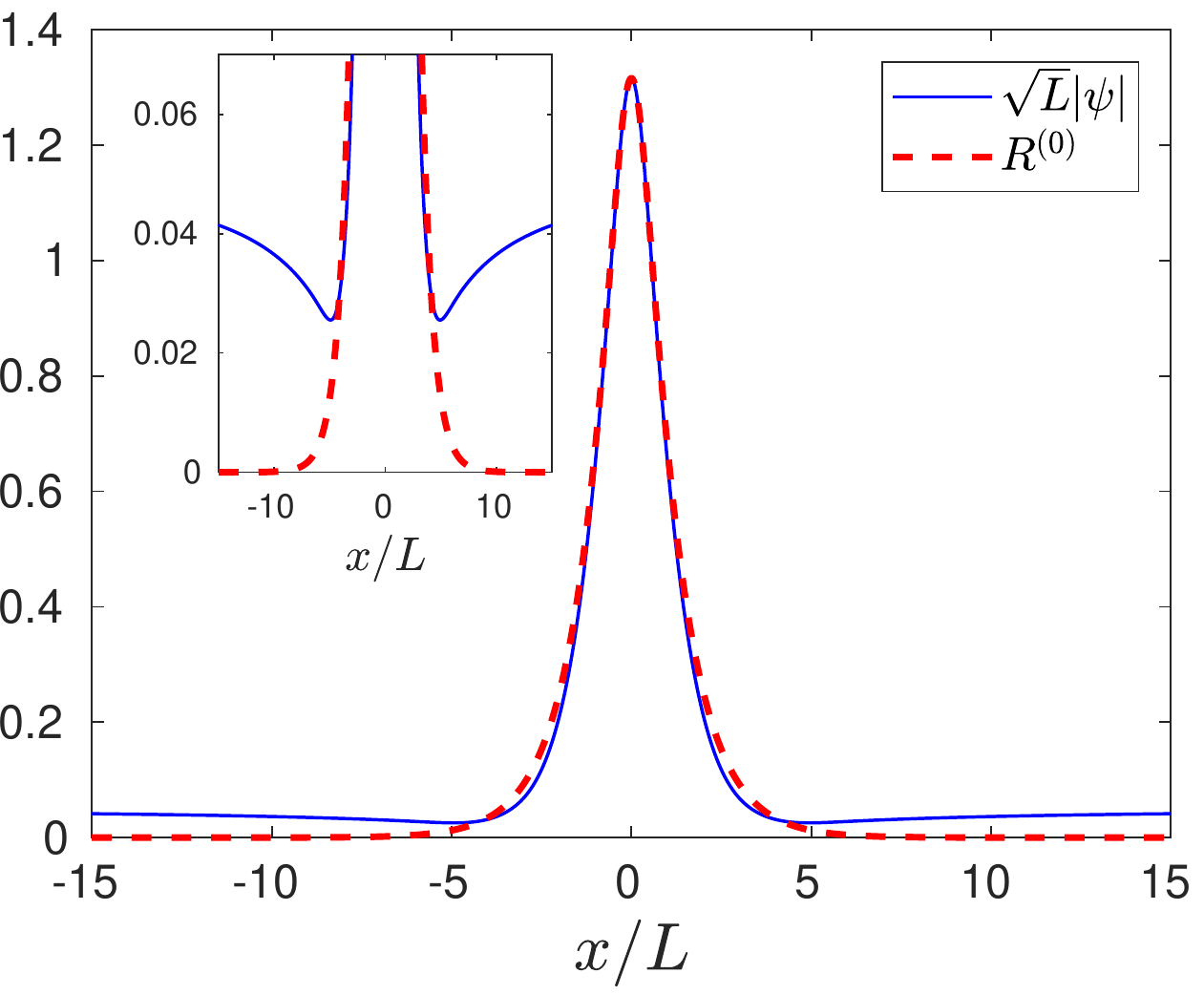}}
	\hspace{0.2cm}
	\subfloat[]{\includegraphics[width=0.32\textwidth]{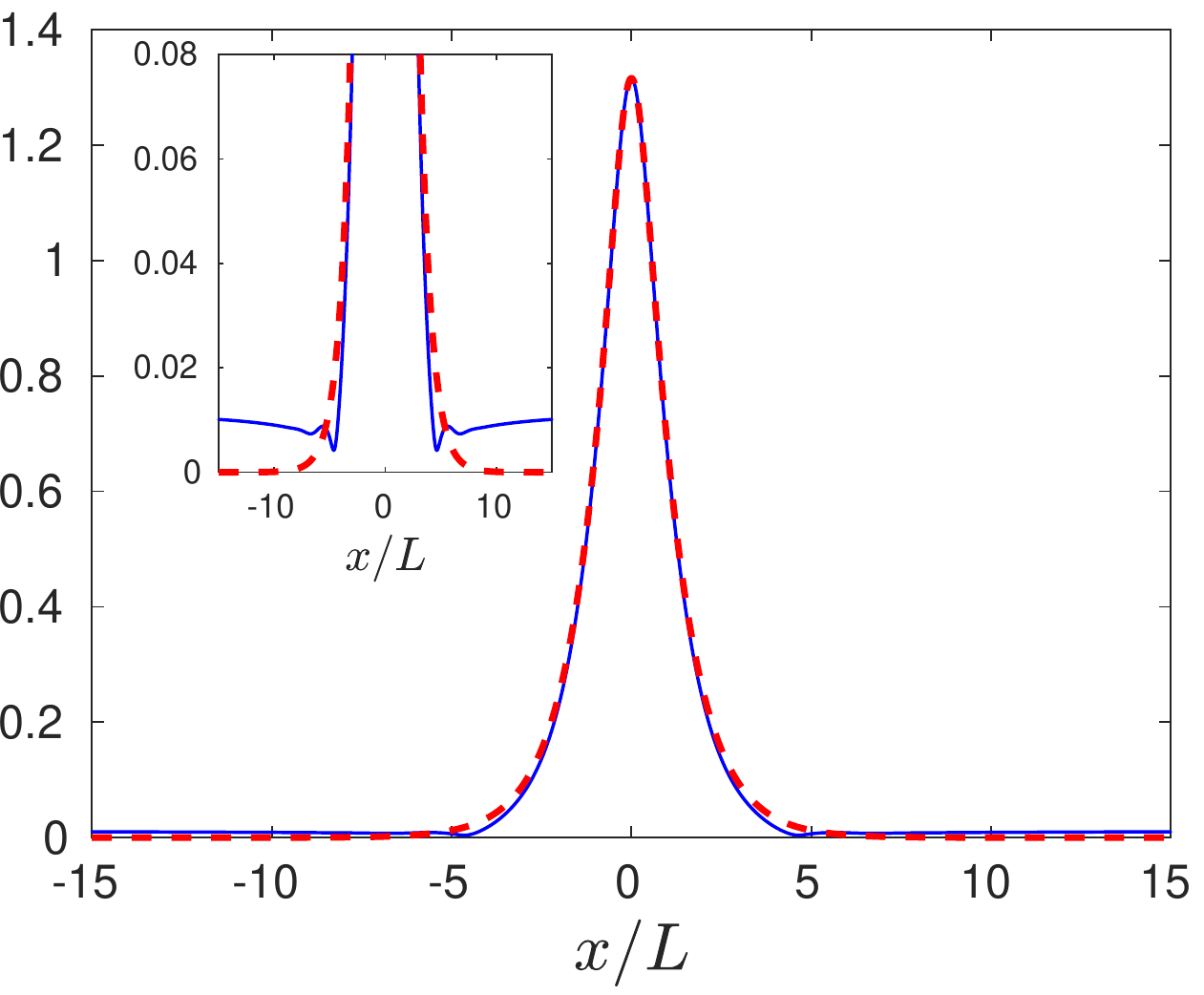}}
	\hspace{0.2cm}
	\subfloat[]{\includegraphics[width=0.31\textwidth]{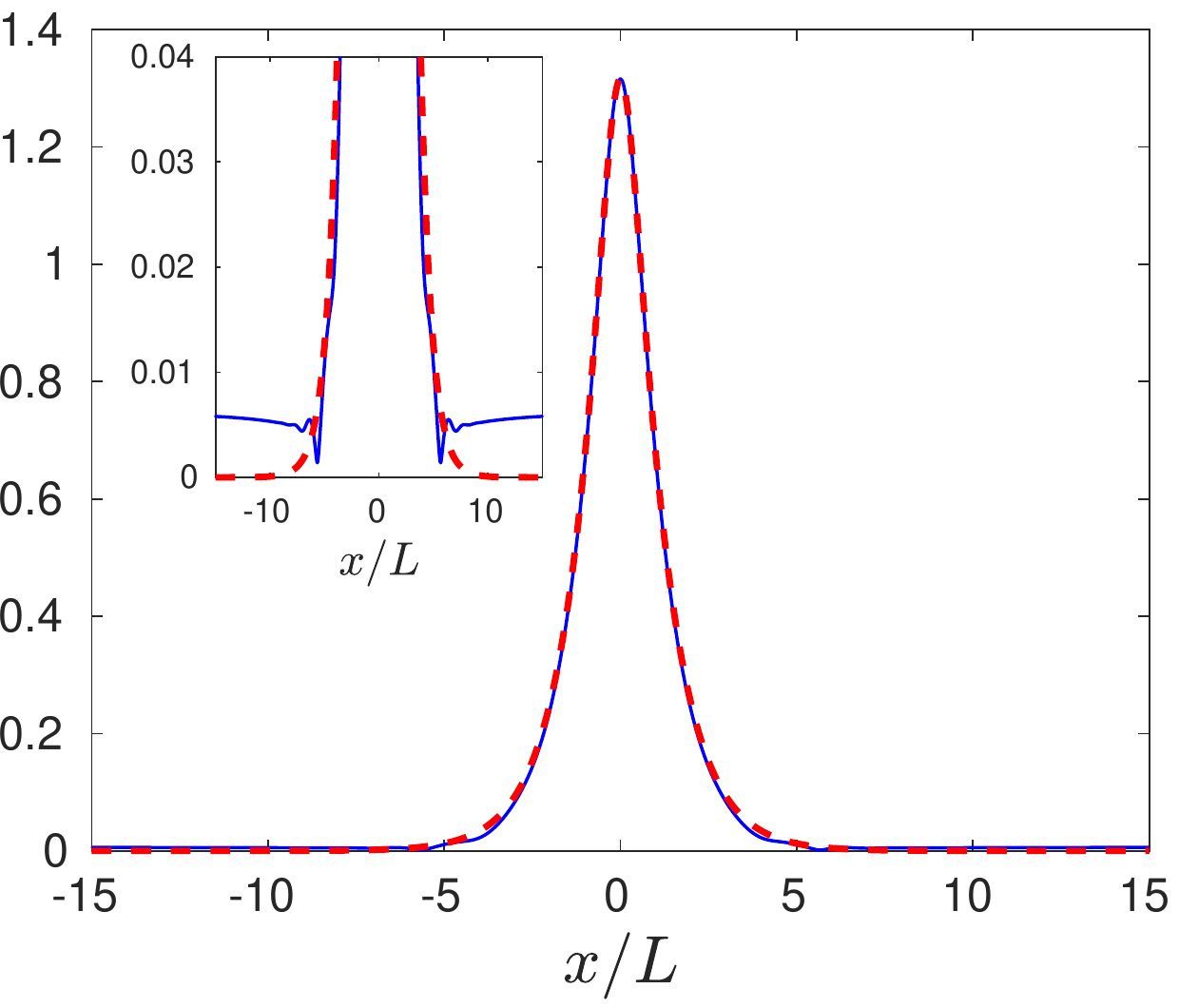}}
	\hspace{0.2cm}
	\subfloat[]{\includegraphics[width=0.34\textwidth]{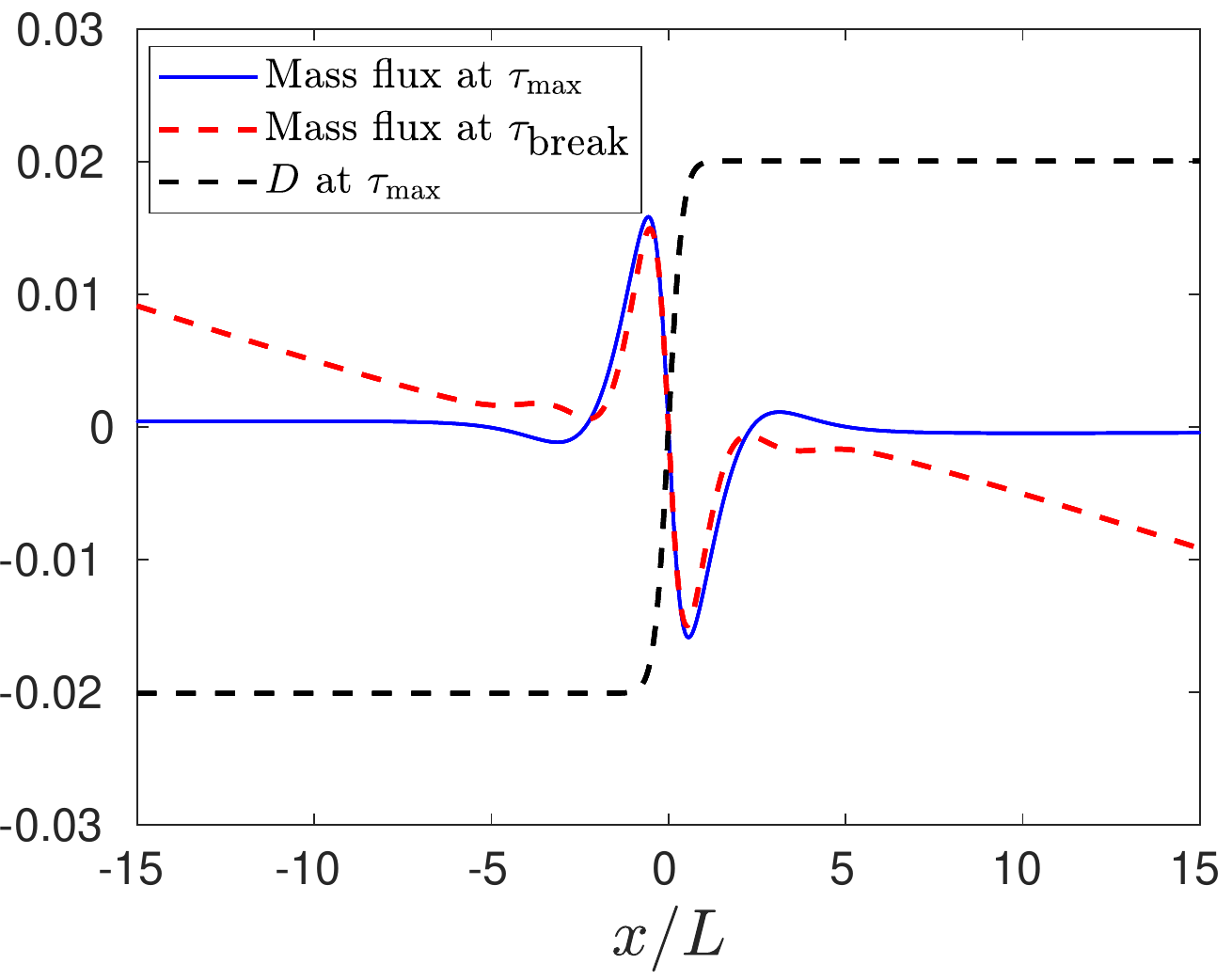}}
 \subfloat[]{\includegraphics[width=0.33\textwidth]{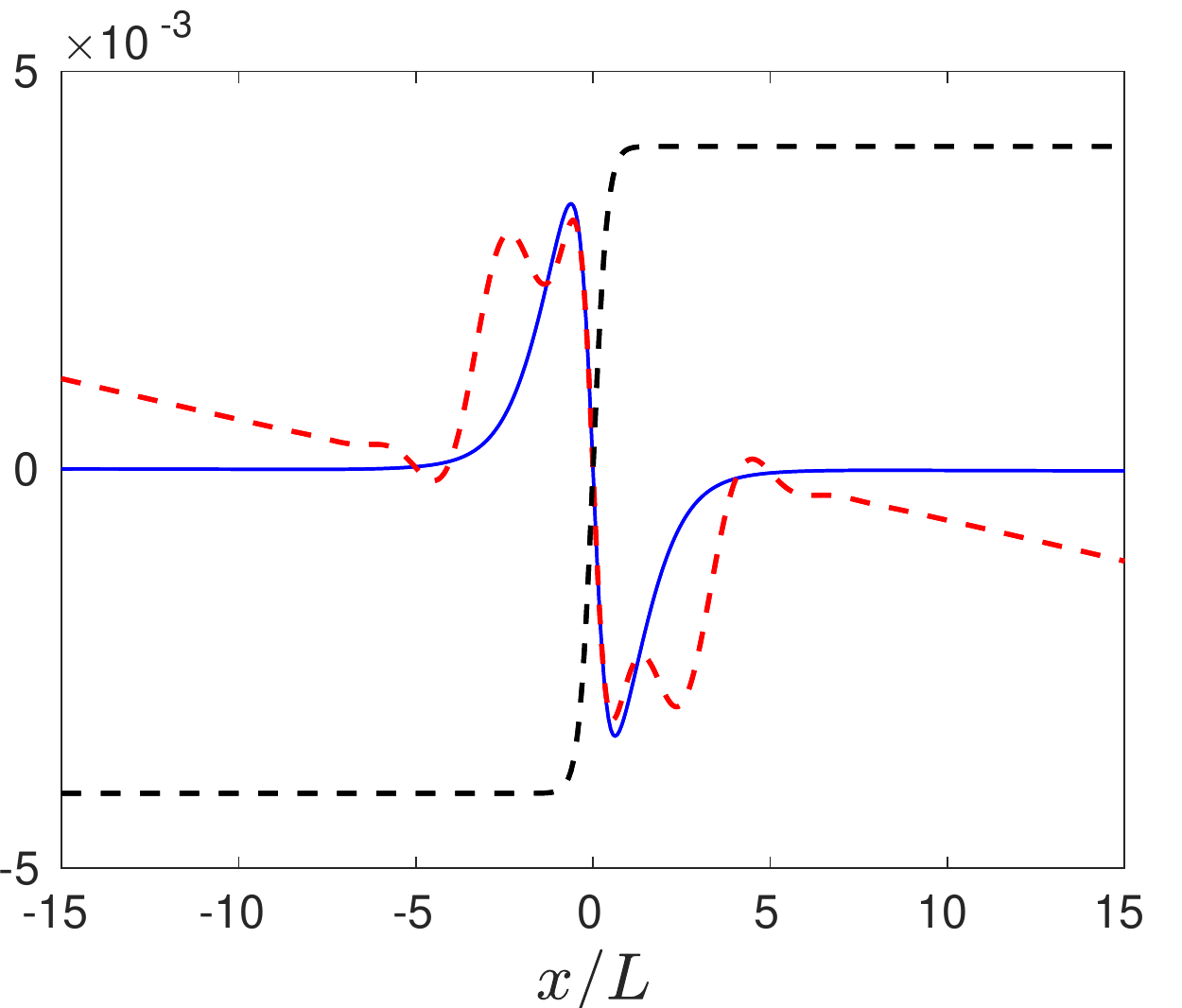}}
	\hspace{0.2cm}
 \subfloat[]{\includegraphics[width=0.31\textwidth]{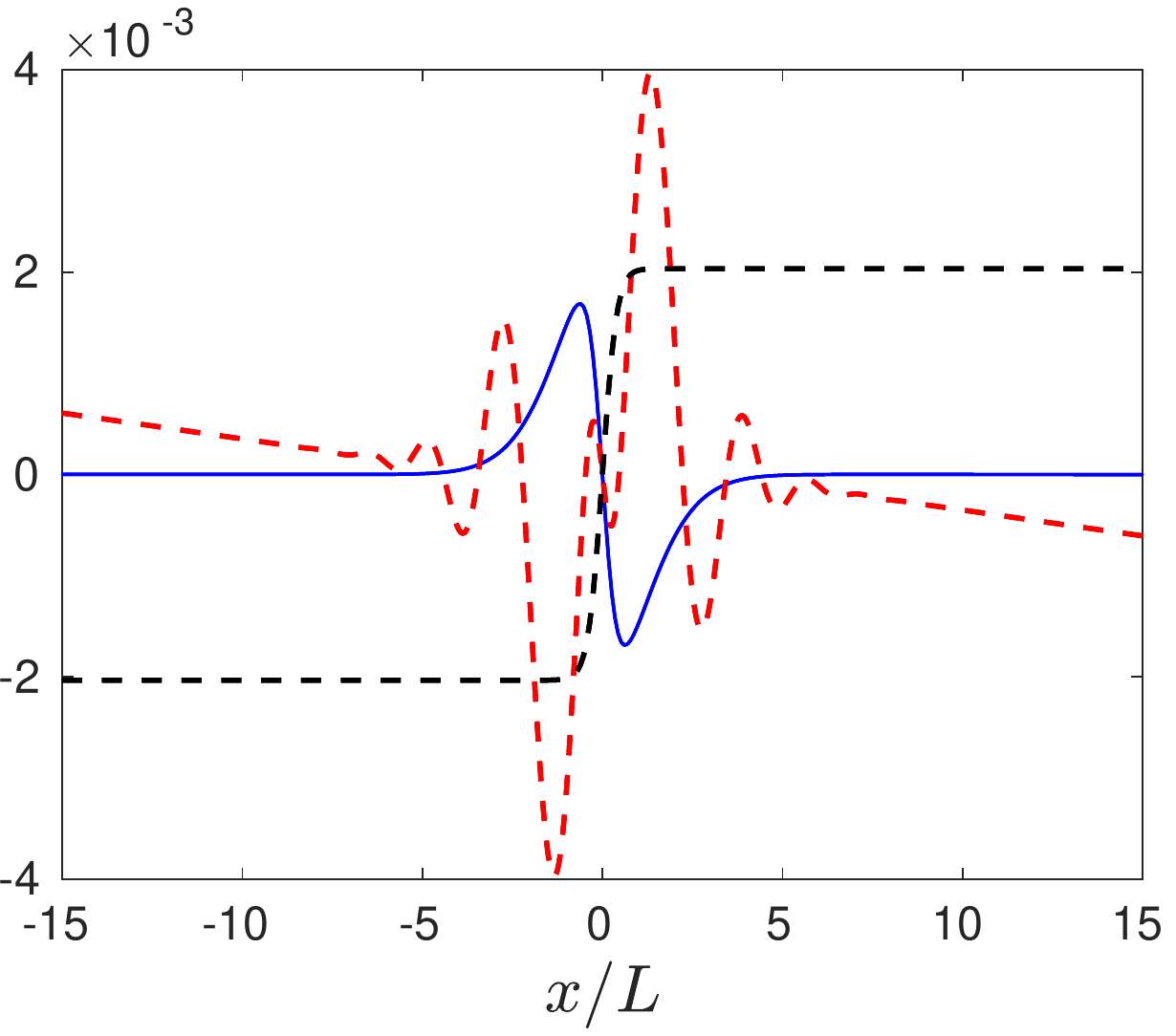}}
	\hspace{0.2cm}
	\caption{Panels (a-c) display the rescaled profile $\sqrt{L}|\psi(x/L,t)|$ vs $R^{(0)}(x)$ at $t_{\textrm{break}}$ for $\delta = 5 \times 10^{-3}$,  $\delta = 10^{-3}$ and $\delta = 5 \times 10^{-4}$. These graphs verify the universal profile  at the breakdown time $t_{\textrm{break}}.$ Insets reveal the formation of small oscillations around the bottom of the solution. In panels (d-f), we displayed the mass flux $J(\xi,\tau)$ and the dissipation term $D(\xi,\tau)$  for the corresponding values of $\delta$. One observes that at $t_{\textrm{break}}$ the influx of mass becomes comparable with dissipation.}
\label{Fig5}
\end{figure}

Adiabatic dynamics takes place in the renormalized wavefunction $\Psi(\xi,\tau)$, see equation (\ref{eq3_2}). In the adiabatic stage,  the interaction (mass transfer) between the collapsing core and the non-collapsing tail is neglected. Since adiabatic approximation breaks down, the first thing that one can check is this mass transfer. The mass balance equation in the renormalized variables for the interval $[-\xi_0,\xi_0]$ is 
\begin{eqnarray}
 \frac{d}{d\tau} \int_{-\xi_0}^{\xi_0}|\Psi(\xi,\tau)|^{2} d\xi   = -\left.J(\xi,\tau)\right\vert_{-\xi_0}^{\xi_0} -\left.D(\xi,\tau)\right\vert_{-\xi_0}^{\xi_0},
\end{eqnarray}
 where $J(\xi,\tau) = \frac{1}{i}\left(\Psi^{*}\Psi_{\xi} - \Psi^{*}_{\xi}\Psi \right)$ and $D(\xi,\tau) = 2\delta\int_{0}^{\xi}|\Psi(\xi^{\prime},\tau)|^{6} d\xi^{\prime}$ are the mass flux and  dissipation term respectively. This relation can be derived using the equation (\ref{eq3_2}). Figure \ref{Fig5} shows the mass flux $J(\xi,\tau)$ and the dissipation $D(\xi,\tau),$ as well as the profile of the solution at $\tau_{\textrm{break}}$ for $\delta = 5 \times 10^{-3},$ $\delta = 10^{-3}$ and $\delta = 5 \times 10^{-4}$. The panels (a-c)  show that at the breakdown instant 

 \noindent the inner core of the solution approaches the universal profile  $R^{(0)}$, which is accompanied by the formation of small oscillations around the bottom of the profile, as can be observed in the corresponding insets. 
 
 Although these oscillations are tiny, the corresponding fluxes of mass may be large due to small wavelengths. Indeed, the comparison between $J(\xi,\tau)$  and $D(\xi,\tau)$ is showed in the Figures \ref{Fig5}(d-f). These figures indicate that both terms, mass flux $J(\xi,\tau)$  and dissipation $D(\xi,\tau)$, become comparable at  $\tau_{\textrm{break}}$. Recalling that adiabatic theory assumes a dominant dissipative effect (mass transfer is neglected),
our \emph{hypothesis} is that the invalidity of the adiabatic approximation is due to this increasing influx of mass into the inner core after the arrest of the collapse. The  small oscillations observed in the insets of Figures \ref{Fig5}(a-c) is the result of this  core-tail interaction.

\begin{figure}[H]
	\centering
	\subfloat[]{\includegraphics[width=0.45\textwidth]{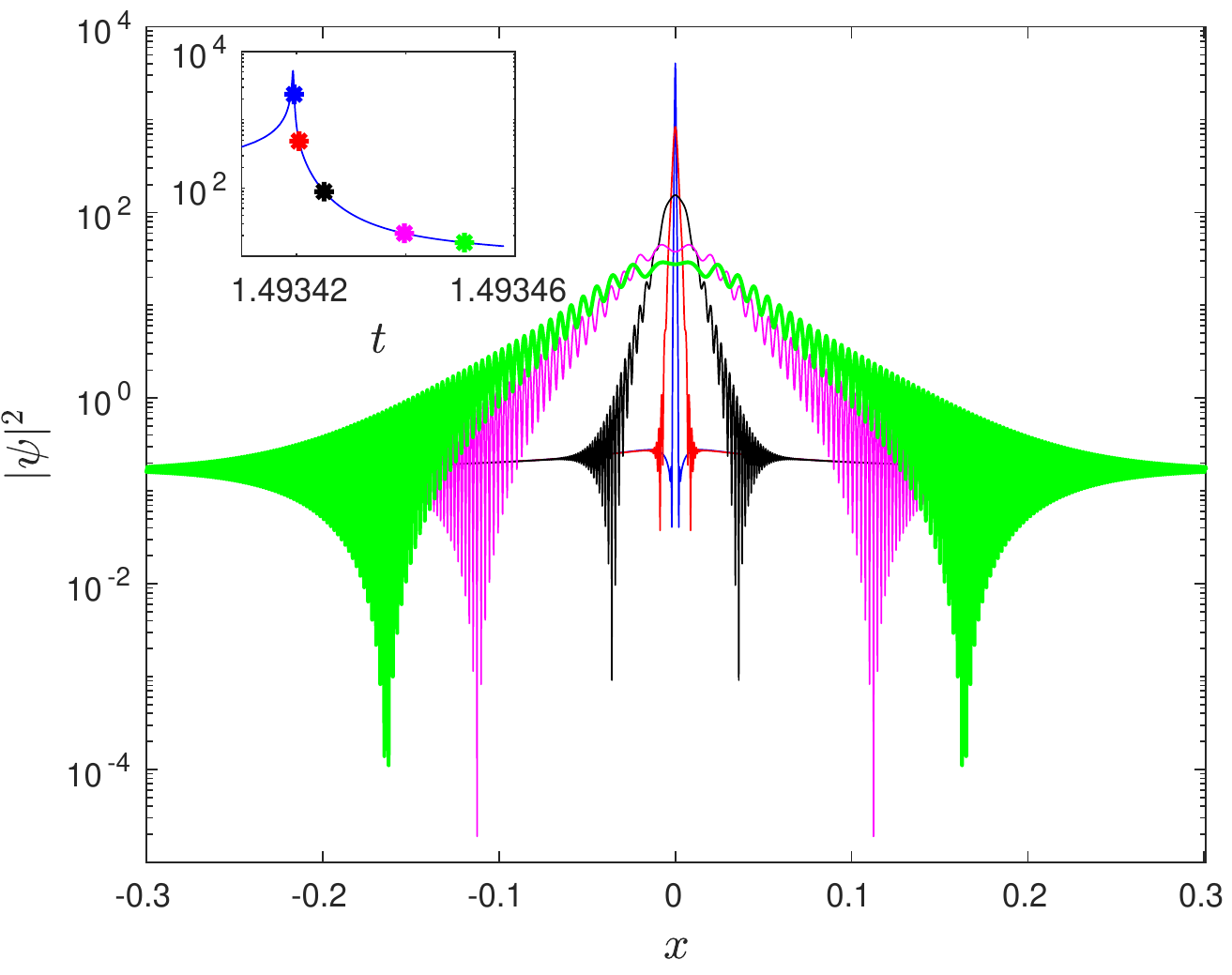}}
	\hspace{0.2cm}
	\subfloat[]{\includegraphics[width=0.45\textwidth]{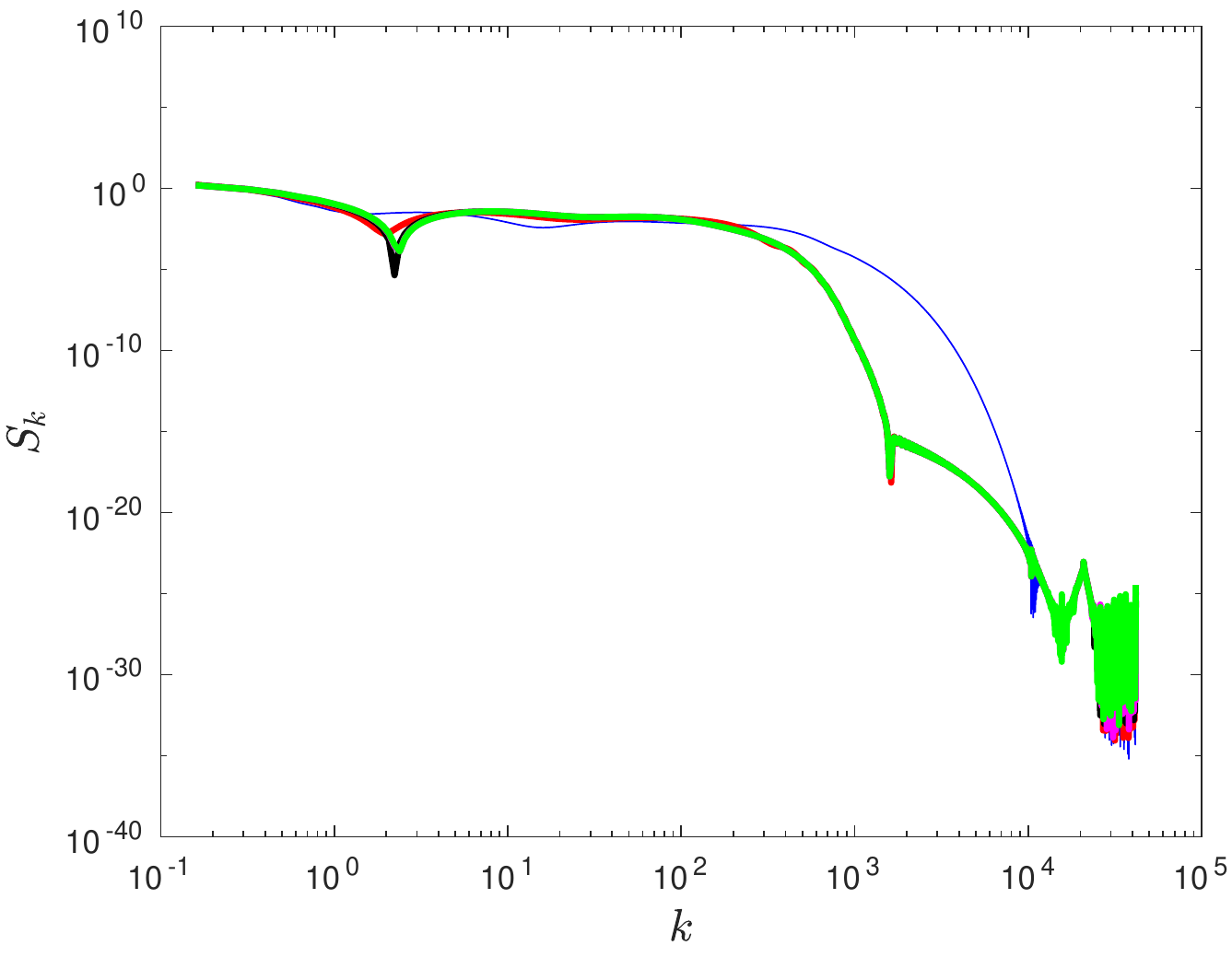}}
	\hspace{0.2cm}
	\caption{(a) Profile of $|\psi|^2$ for $\delta = 10^{-3}$ at different instants of time in the post-adiabatic dynamics. The first blue profile corresponds to the time $t_{\textrm{break}}$. The high-frequency oscillations  propagate in both directions spreading in the whole collapsing core. The inset shows the function $1/L(t)$ and the corresponding  instants considered in post-adiabatic dynamics. (b) Spectra of the solution from the left panel, which almost collapse except at the earlier time. These quasi unmodified spectra support that the dynamics is almost linear. }
\label{Fig7}
\end{figure}

\section{Post-adiabatic dynamics }

The goal of this section is to describe the post-blowup dynamics at instants  after the breakdown of the adiabatic stage. We will show an evidences that after the end of the adiabatic phase a quasi-linear regime is developed.  We also observe  that the ansatz (\ref{eq3_10}) still holds approximately in such quasi-linear regime. 

\subsection{Quasi-linear regime} 

Numerical simulations for the damped NLS (\ref{eq1}) indicate the generation of small oscillations around the collapsing core short after the invalidity of the adiabatic stage. When the solution defocuses, the amplitude and the number of these outgoing fluctuations increase and  ``pollute" the inner core of the  solution. Notice that the oscillations appear for the absolute value $|\psi|^2$, and we will see  that they result from an interference of the linear wave with the tail. In Figure  \ref{Fig7}(a) we have plotted some snapshots of $|\psi|^2$ for $\delta = 10^{-3}$ at different instants of the defocusing process, starting  at the breakdown time $t_{\textrm{break}}$. In the  corresponding inset, the function $1/L(t)$ and the 

\begin{figure}[H]
	\centering
	\subfloat[]{\includegraphics[width=0.4\textwidth]{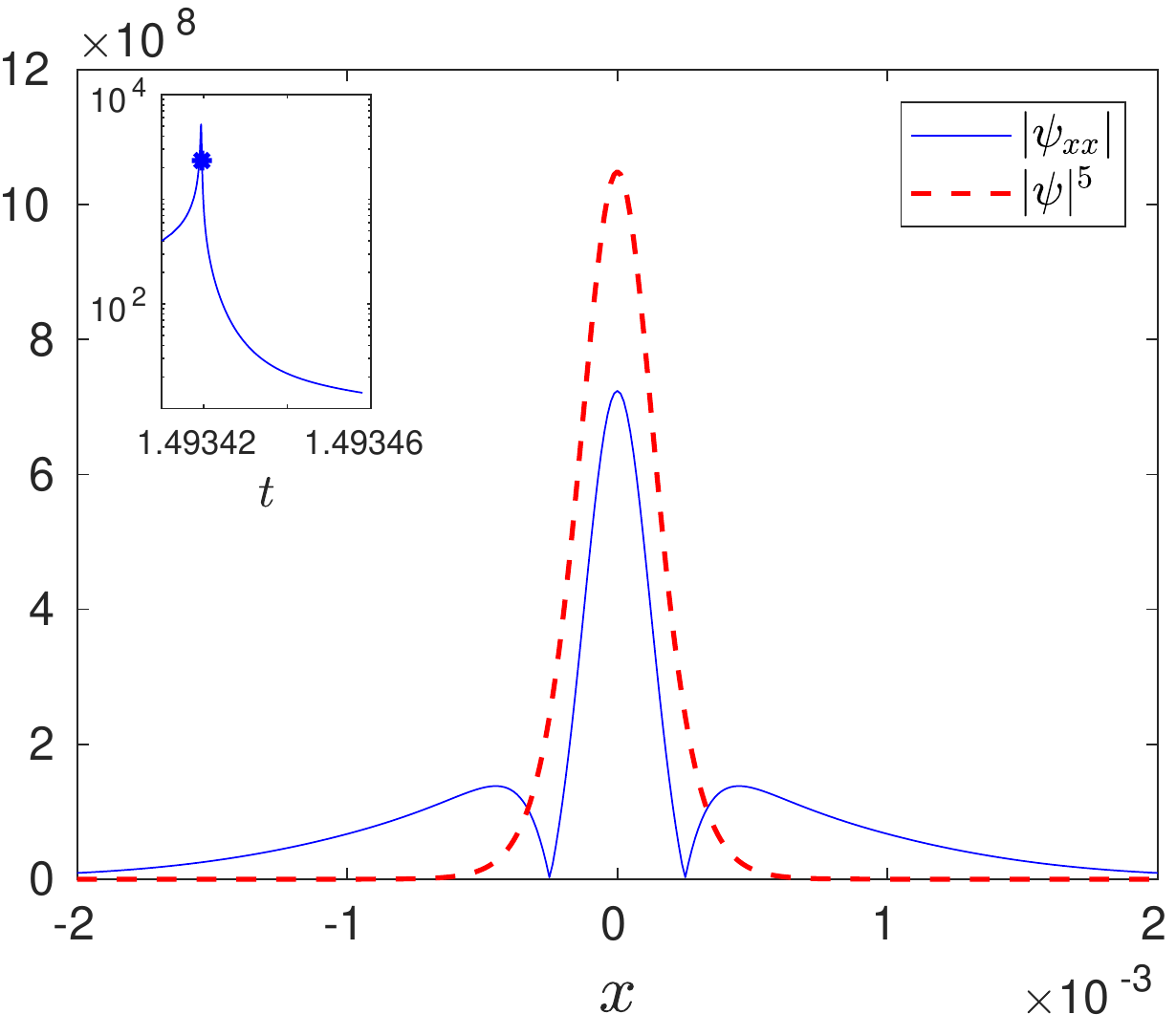}}
	\hspace{0.2cm}
	\subfloat[]{\includegraphics[width=0.42\textwidth]{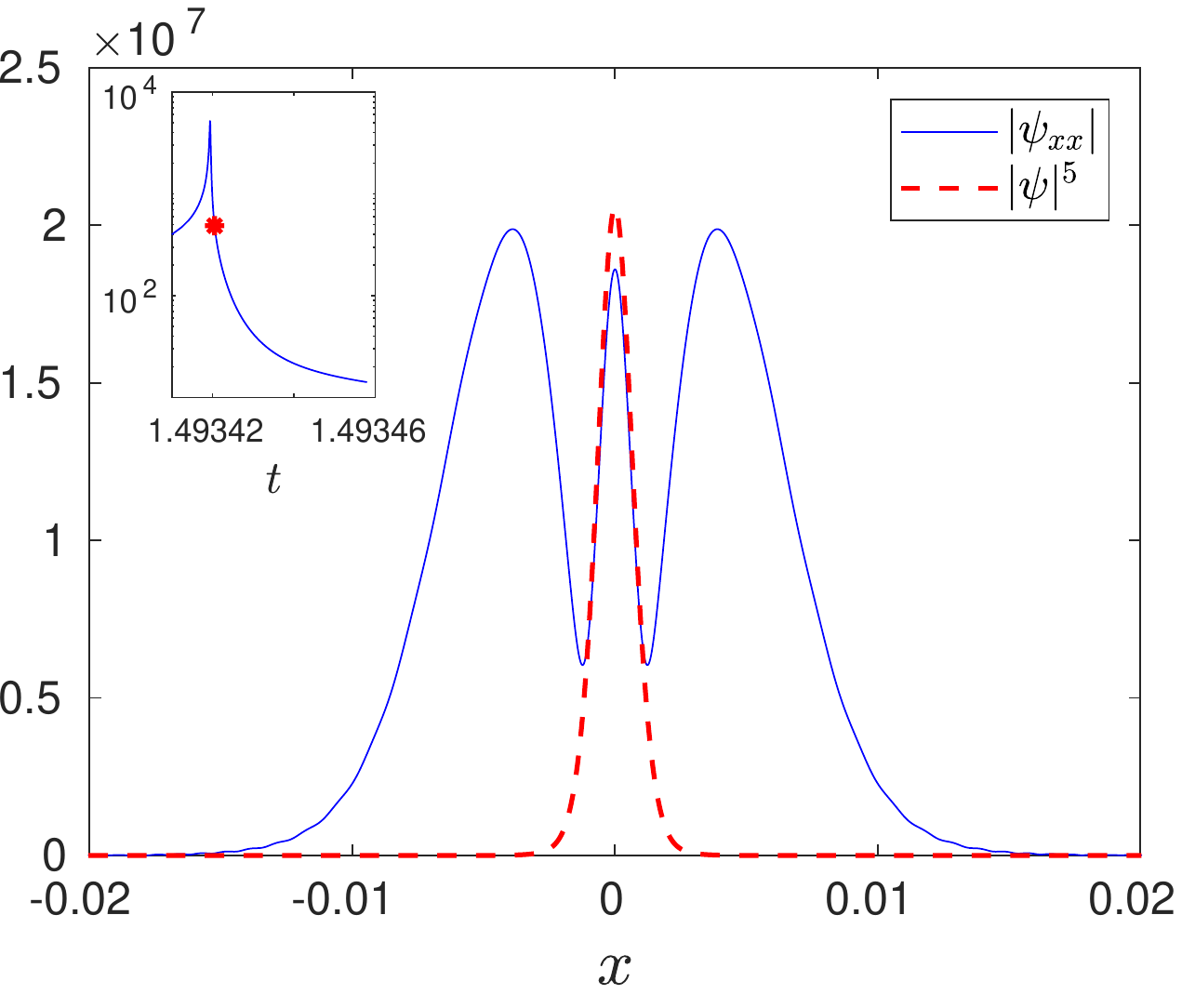}}
	\hspace{0.2cm}
 \subfloat[]{\includegraphics[width=0.42\textwidth]{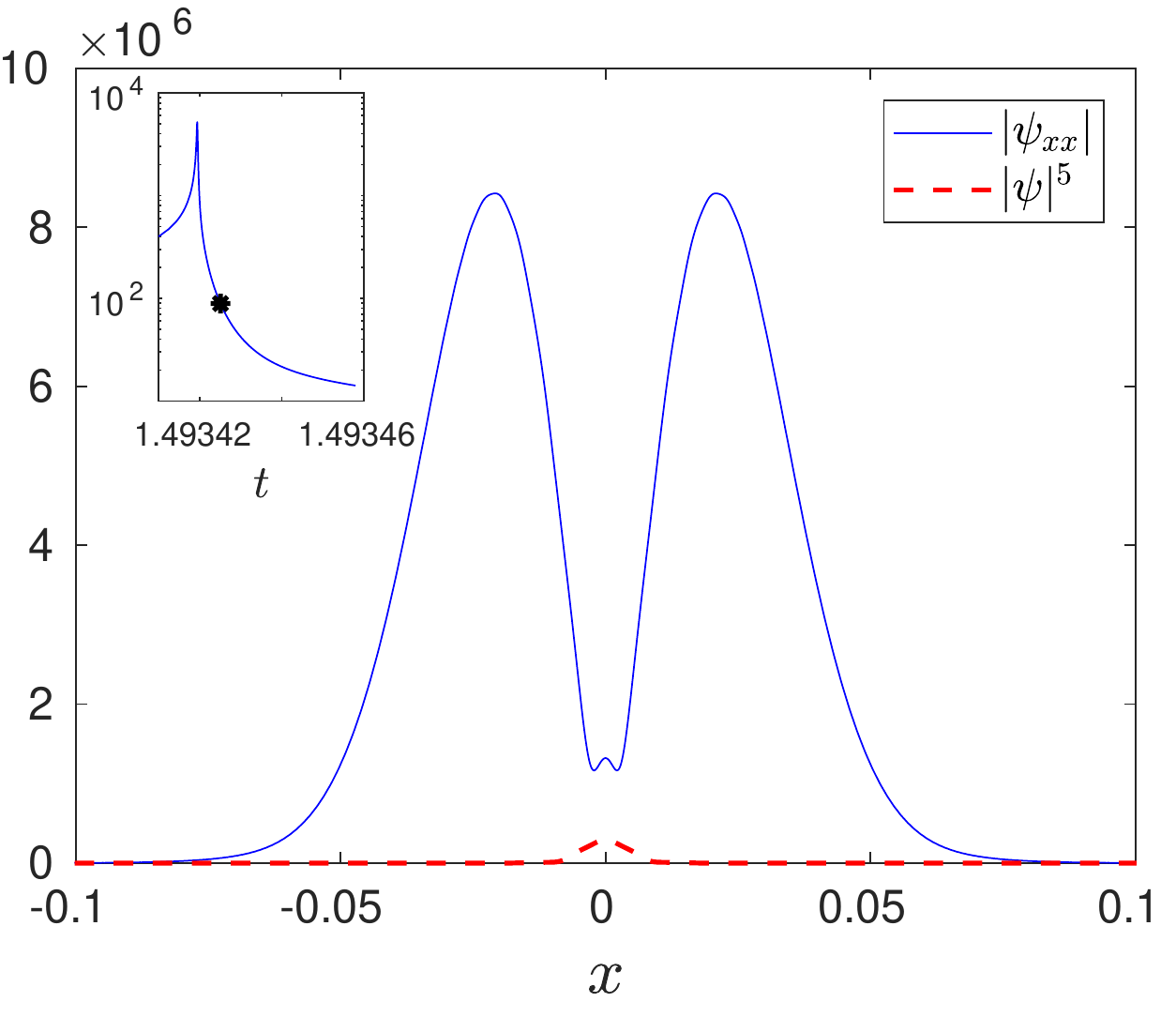}}
	\hspace{0.2cm}
 \subfloat[]{\includegraphics[width=0.4\textwidth]{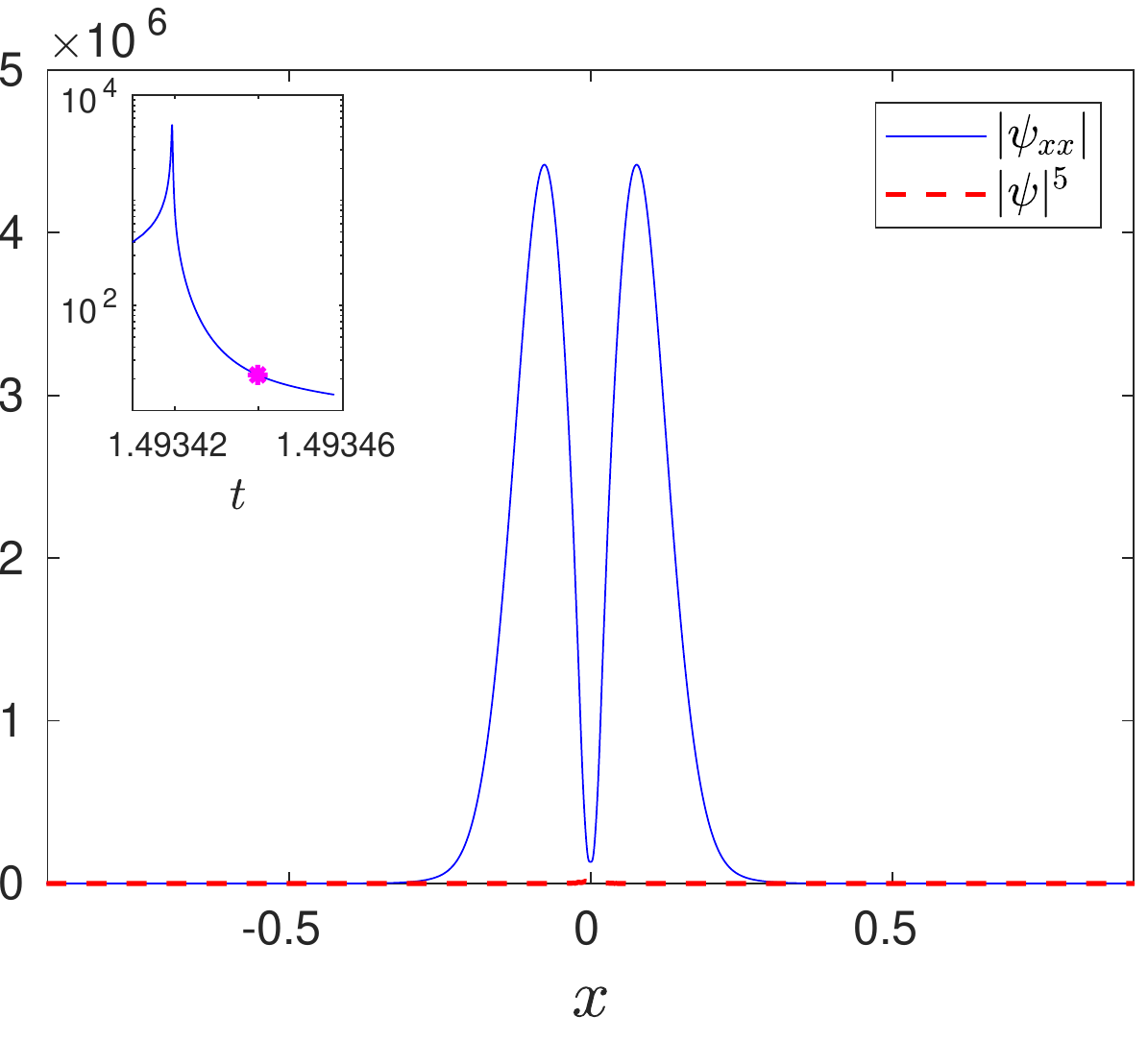}}
	\hspace{0.2cm}
 \subfloat[]{\includegraphics[width=0.4\textwidth]{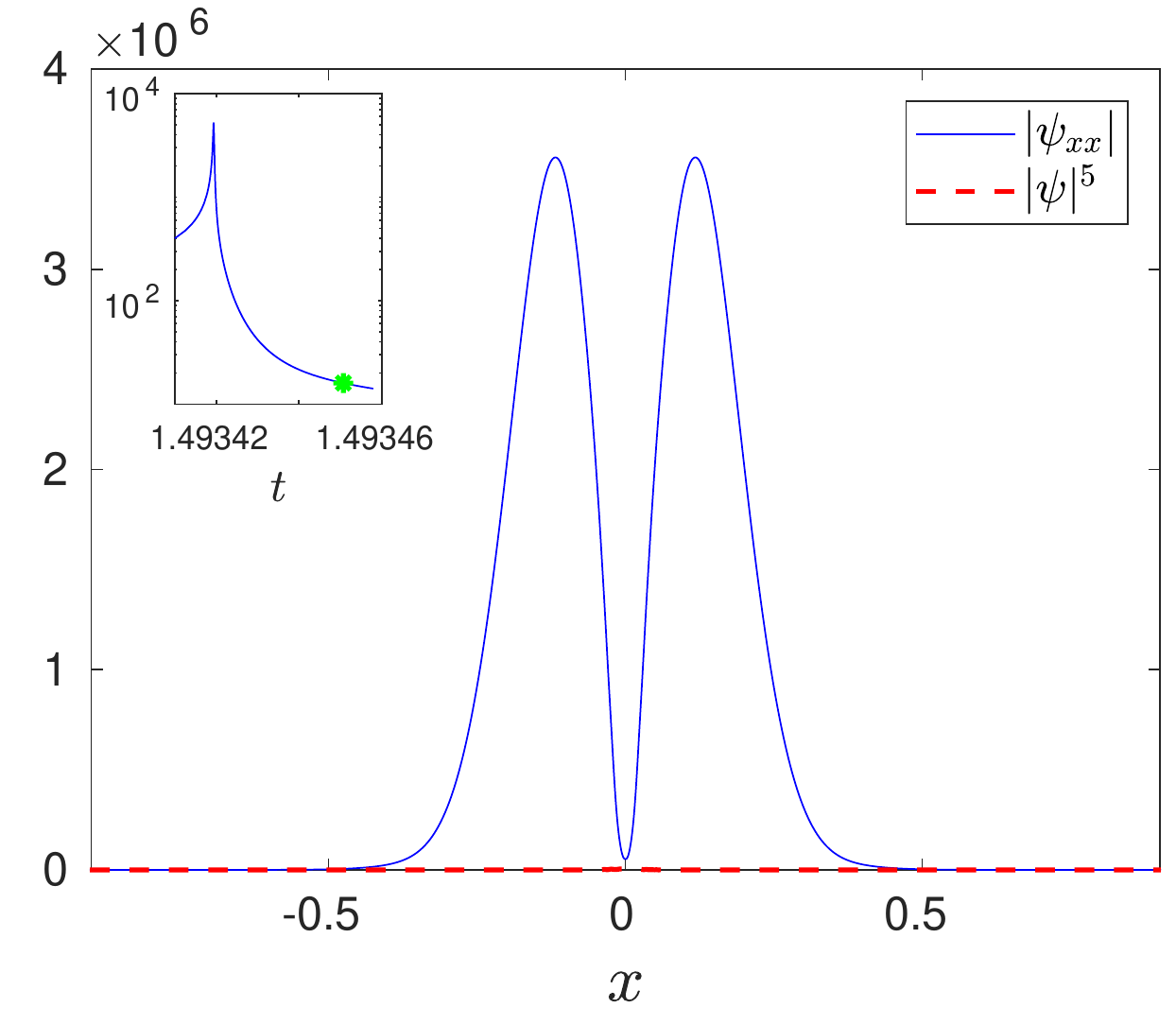}}
	\hspace{0.2cm}
	\caption{Comparison between the dispersive term $|\psi_{xx}|$ and the nonlinear term $|\psi|^5$ for $\delta = 10^{-3}$ at different moments after the breakdown of the adiabatic stage. As time runs, the  balance between both terms is replaced by a dispersive-dominated stage caused by the rapid defocusing process. Insets indicate the time of each plot.}
\label{Fig8}
\end{figure}
\noindent times considered are displayed. Here the function $L(t)$ was computed by using  the relation (\ref{eq_L}).

The generation of these oscillations strongly suggests a dispersive-dominated regime in the post-adiabatic dynamics. In Figure \ref{Fig7}(b) we plotted the absolute value of the spectrum of the solution, $S_{k} = |\hat{\psi}(k)|^2 + |\hat{\psi}(-k)|^2,$ where $\hat{\psi}(k) = \mathcal{F}[\psi](k)$ is the Fourier transform. In this figure, one can see the almost unchanged spectra of the solution in the prescribed regime: the spectrum corresponding to the red, black, pink and green instants are almost identical, but different from the spectrum at the blue instant near the peak. It indicates a quasi-linear regime in which the dispersive term  dominates the nonlinear one. This quasi-linear regime was also reported in \cite{Passot05}, by a numerical study of the two-dimensional critical NLS with a supercritical nonlinear damping.

In order to reinforce the evidences of this quasi-linear regime, we complement the previous indicia by checking directly the evolution of the dispersive and nonlinear terms. In Figure \ref{Fig8} we have compared the terms $|\psi_{xx}|$ and $|\psi|^5.$ At first instants in the defocusing process both terms have the same order of magnitude, as expected. Afterwhile, $|\psi_{xx}|$ becomes dominant over $|\psi|^5$, in such a way that the space-interval in which the former  strongly  dominates the latter expands in time. Despite both terms decrease when the solution defocuses, the fast decay of the nonlinear term induces a quasi-linear dynamics. Consequently, in that stage the resulting dynamics of the solution is approximately described by the linear Schr\"odinger equation. 

 \subsection{Persistence of the universal profile} 
 
 So far, we have observed two stages in the post-blowup dynamics: an adiabatic regime, in which the inner core of the solution follows asymptotically the universal profile (\ref{eq3_10}) subjected to the reduced equations (\ref{eq3_11}), and a quasi-linear stage,  caused by the fast decay of the peak of the solution. In such quasi-linear regime reduced equations are not valid. Now, our goal is to show that (\ref{eq3_10}) remains valid in the linear regime. 

\begin{figure}[H]
	\centering
	\subfloat[]{\includegraphics[width=0.39\textwidth]{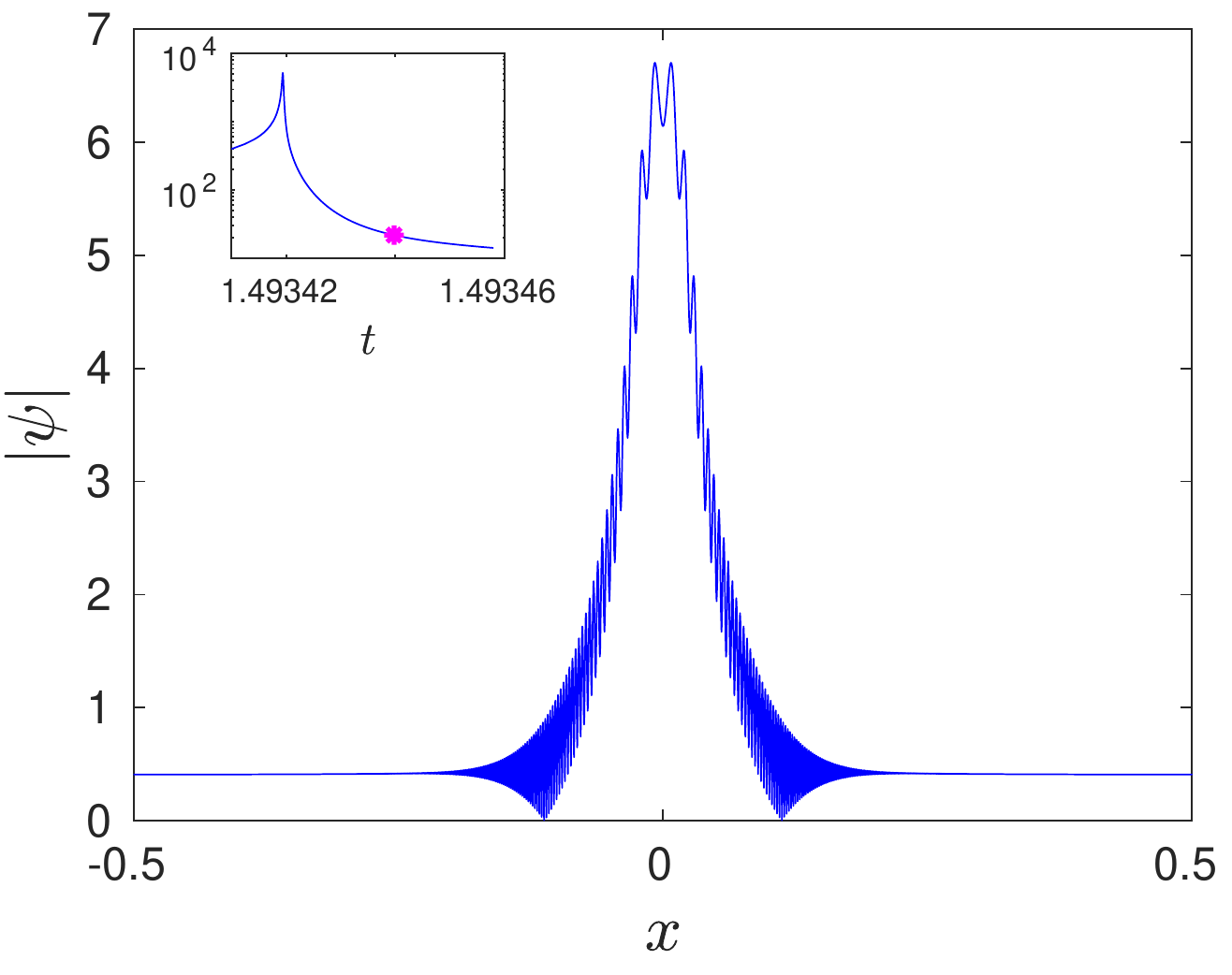}}
	\hspace{0.2cm}
	\subfloat[]{\includegraphics[width=0.35\textwidth]{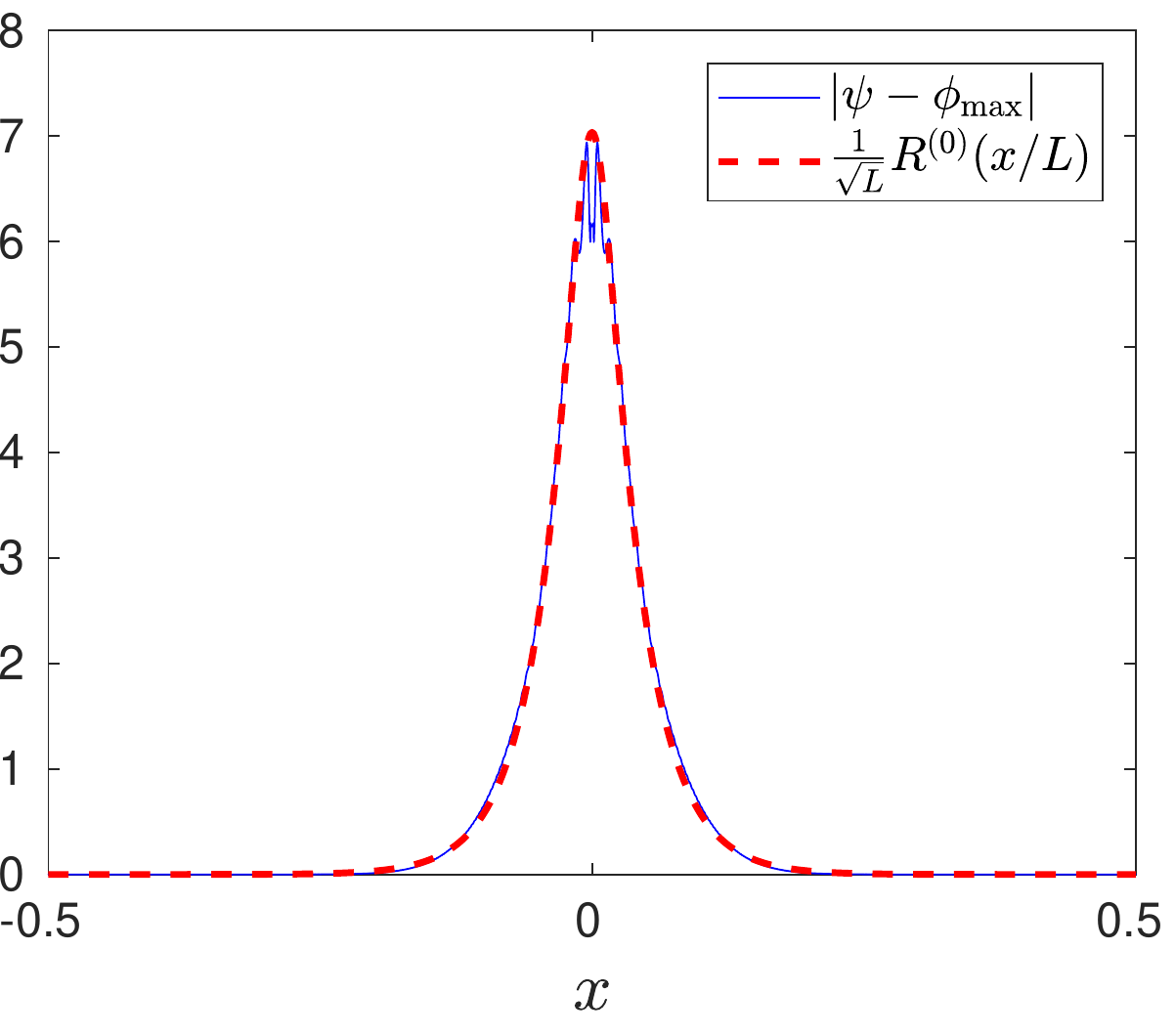}}
	\hspace{0.2cm}
 \subfloat[]{\includegraphics[width=0.4\textwidth]{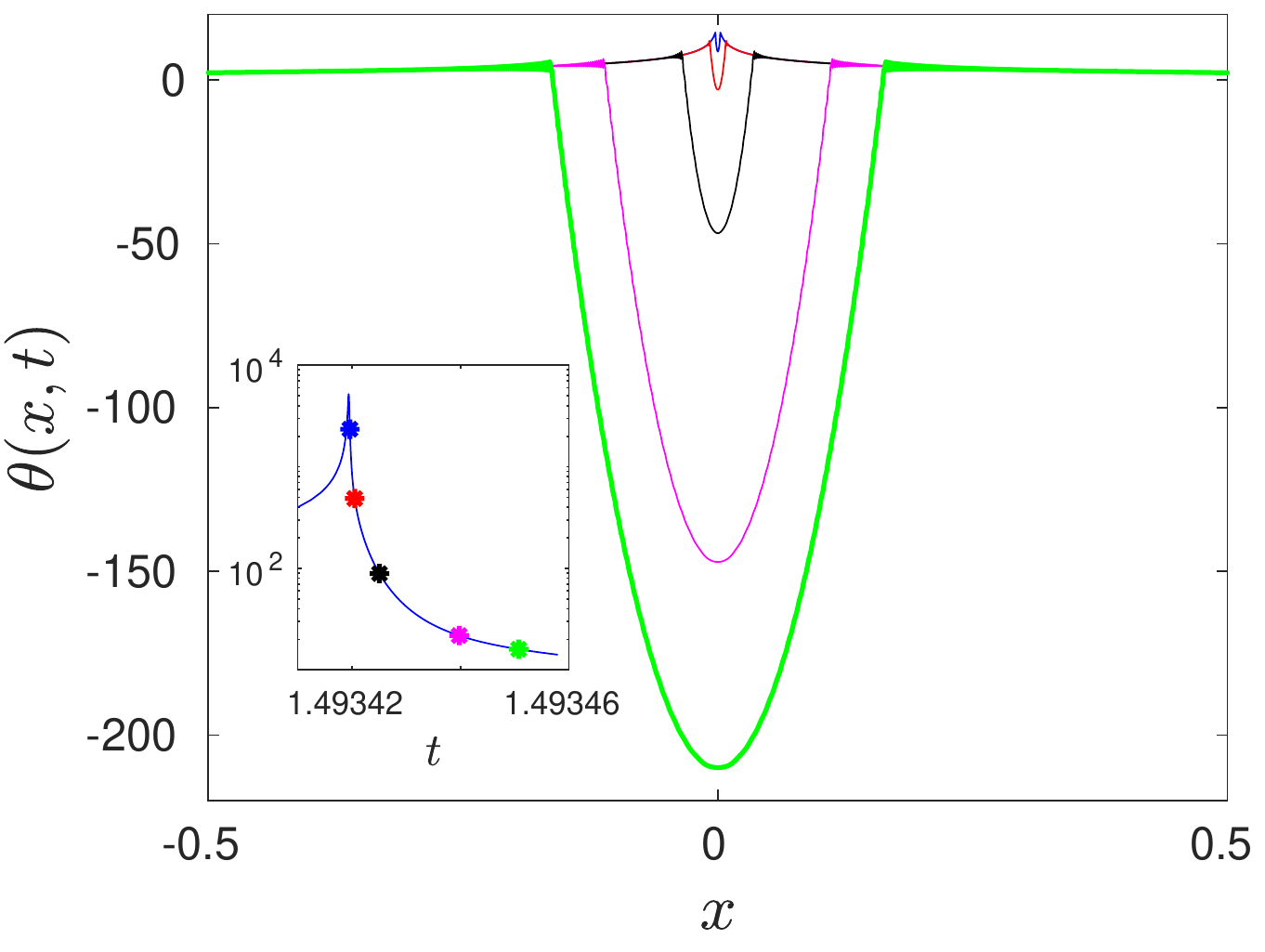}}
	\hspace{0.2cm}
 \subfloat[]{\includegraphics[width=0.35\textwidth]{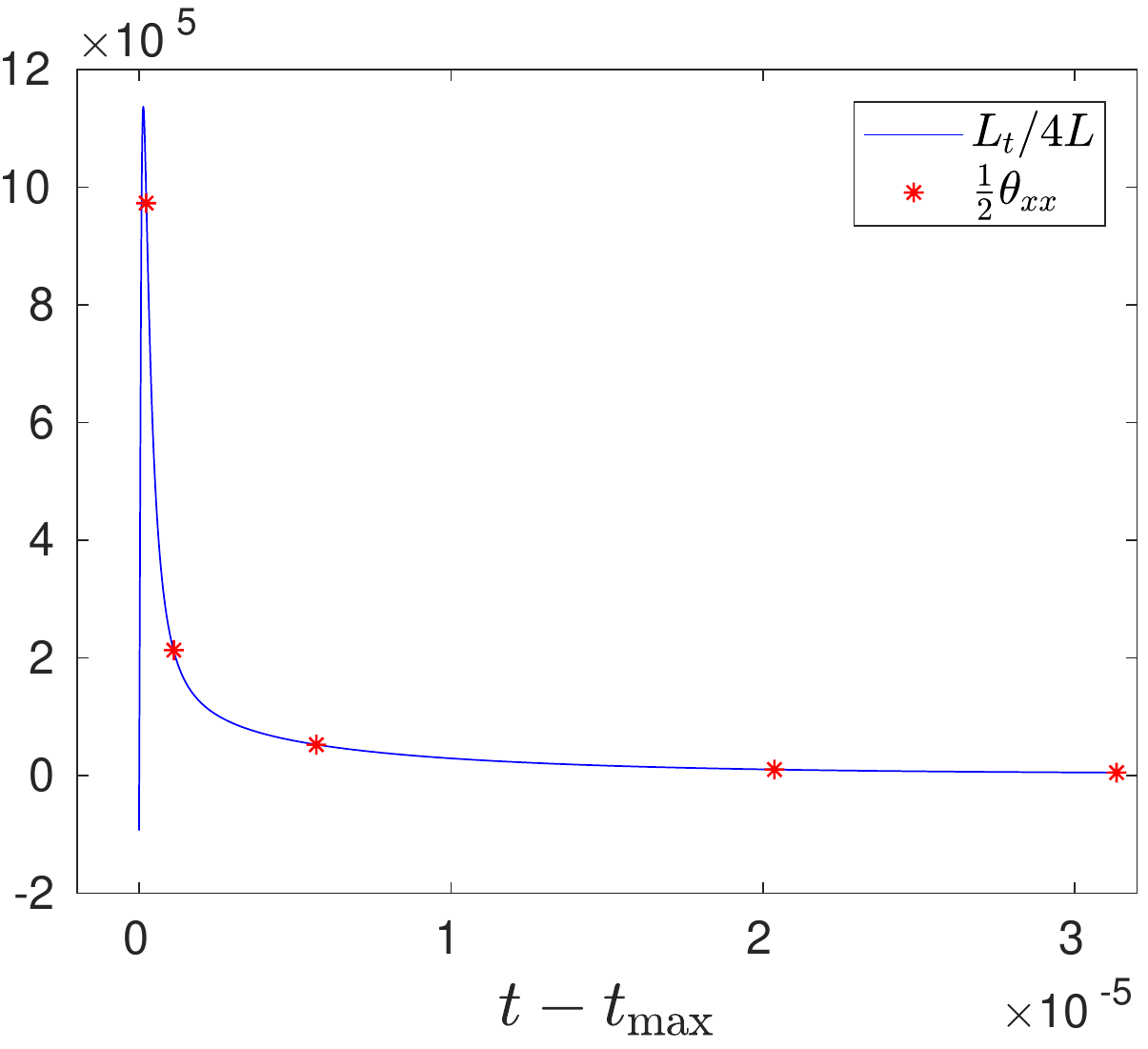}}
	\hspace{0.2cm}
	\caption{Verification of the universal profile in the quasi-linear regime of the post-blowup dynamics. (a) Profile of $|\psi|$ for $\delta = 10^{-3}$ at certain moment in the quasi-linear stage. In the corresponding inset indicates the time of the plot. (b) Comparison between the cleaned profile $|\psi-\phi_{\max}|,$  and the rescaled ground state $R^{(0)}.$ Then, the  collapsing core shape is modulated by $R^{(0)}.$ (c) Phase of the solution $\theta(x,t)$ at different moments in the post-adiabatic dynamics. Insets show the time of the plot. (d) Leading coefficients of the quadratic functions, $\frac{1}{2}\theta_{xx}$, plotted on the graph of $L_{t}/4L$ with $L(t)$ computed by the relation (\ref{eq_L}).}
\label{Fig9}
\end{figure}

Due to the interference between the inner core and the tail in the quasi-linear regime, collapsing core is typically ``polluted" by small oscillations, see Figure \ref{Fig9}(a). Therefore, in order to  verify (\ref{eq3_10}), we need to ``clean up'' the profile of $|\psi|$ by removing almost all the extra mass (mass above $M_c$) contained in the solution, and compare this cleaned profile  with the rescaled ground state. The extra mass contained in the solution, $\phi_{\max}(x)$, can be approximated by cutting the inner zone of 

\noindent the solution off  at the instant of maximum amplitude, i.e.,
\begin{eqnarray}
    \label{94}
     \phi_{\max}(x) =\left\{\begin{array}{ll}
         \psi(x,t_{\max}),& |x|\geq 6L(t_{\max}),\\[3pt]
            0,& |x| < 6L(t_{\max}).
            \end{array}\right.
\end{eqnarray}
Then, the cleaned profile is obtained by subtracting from $\psi$ the function $\phi_{\max}(x)$. Indeed, Figure \ref{Fig9}(b) displays  the comparison between $|\psi-\phi_{\max}|$  with the rescaled ground state by using $L \approx 0.035$ for $\delta = 10^{-3}$ in the moment of the linear regime corresponding to Figure \ref{Fig9}(a). The agreement observed verifies the approximation of the collapsing core  by the ground state.

 The next step in the  verification of  the universal profile (\ref{eq3_10}), is to check that the phase of the solution $\theta = \arg \psi(x,t),$ is approximated by  the quadratic expression $\theta(x,t) \approx \tau + \frac{L_t}{4L}x^2.$ In fact, as can be seen in Figure \ref{Fig9}(c), the phase of the solution as function of $x$, through the post-blowup dynamics behaves like a parabola in a vicinity  of $x = 0$. In the bottom row, Figure \ref{Fig9}(d), one can see the good agreement between the leading coefficients of these quadratic functions, $\frac{1}{2}\theta_{xx}$, with the term $L_{t}/4L$ computed by using the relation (\ref{eq_L}). Consequently, the quadratic phase  is verified. 
 
 In conclusion, after removing some oscillations due to the interference with the tail, the collapsing core of the solution  can be approximated  by the universal  profile (\ref{eq3_10}). The validity of (\ref{eq3_10})  in the quasi linear regime was an unexpected finding, because in the collapsing process dispersion and nonlinearity terms are almost balanced, but the numerical simulations suggest that the same ansatz remains valid in the quasi linear regime.

 \subsection{Conjecture: Instantaneous radiation of the critical mass} 
 
\begin{figure}[H]
	\centering
 \includegraphics[width=0.45\textwidth]{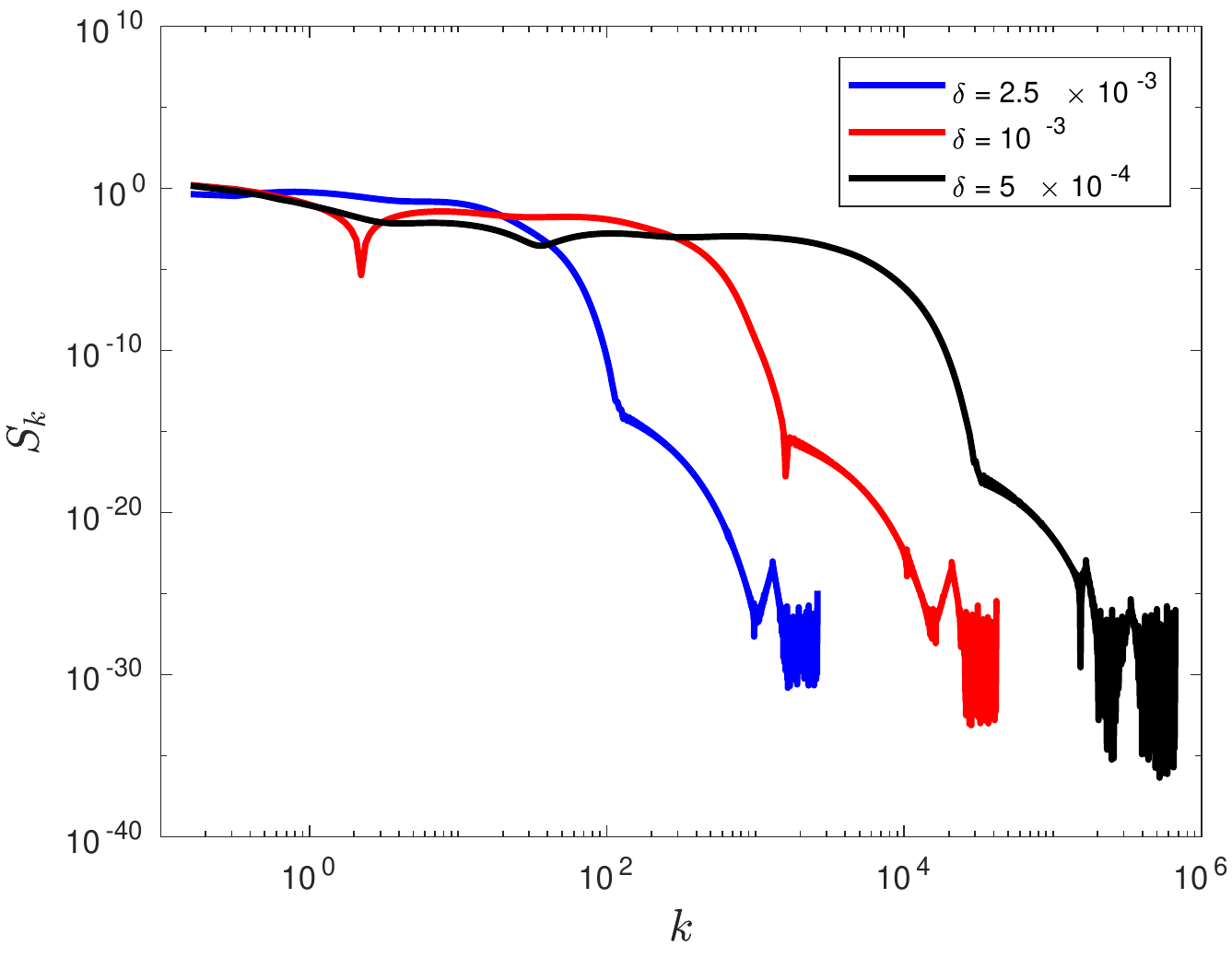}
	\caption{Spectra of the solution for various values of $\delta.$ A tendency to the formation of small scales as $\delta$ decreases is observed. Consequently, our measurements suggest that in the limit of vanishing damping, all the critical mass $M_c$ is instantly radiated to infinity (in the free-space domain) at the blowup time.}
\label{Fig10}
\end{figure}

 In the previous section we observed a quasi-linear regime in the post-adiabatic stage. Due to the increase of defocusing process as $\delta$ decreases, one expects that such regime begins sooner (closer to the peak) and becomes faster in the limit $\delta \to 0$. In this  stage the dynamics can be described by the linear Schr\"odinger equation 
	\begin{eqnarray}
	\label{9.5}
	 i\psi_{t} + \psi_{xx} = 0.
	\end{eqnarray}
	Therefore, the localized part of the solution (collapsing core) spreads out due to the dominant dispersion effect, i.e., each Fourier mode travels at a corresponding wave velocity. As it is  well known, the dispersion relation associated to the equation (\ref{9.5}) is given by $w(k) = k^2.$ Consequently,  the mass of the collapsing core tends to be radiated outward, and therefore towards the domain boundary, through the range of large group velocities $v_{\textrm{g}} = 2k.$ Figure  \ref{Fig7}(a) displays the profile of $|\psi|^2$ at different moments of  the defocusing process for $\delta = 10^{-3}.$  The snapshots clearly show the  radiation of the mass of the collapsing core through the outgoing  waves. In Figure \ref{Fig10} we have plotted the spectrum $S_k$ of the solution  for the three smaller $\delta = 2.5 \times 10^{-3},10^{-3},5 \times 10^{-4}$ at certain instants of the post-adiabatic stage. It indicates that after the arrest of collapse the spectrum exhibits the rapid formation of smaller scales (larger $k$) as long as $\delta$ decreases. Therefore,  we \emph{conjecture} that  in the limit of vanishing dissipation the full collapsed mass $M_c$ is instantly radiated. For example, if the solution is considered in a free-space, $x\in \mathbb{R}$, the mass of the core may instantly radiate to infinity at the collapsing time. 

By virtue of the mass radiation process, the boundary conditions assumed play a crucial role in order to have a well-defined post-blowup dynamics. Probably, in order to have a better description of the dynamics  a kind of absorbing boundary conditions should be used in physical applications and numerical simulations.

 \section{Conclusion}

In this work we have provided a systematic numerical study of the post-blowup dynamics of singular solutions in the framework of the one-dimensional critical focusing  NLS equation with a small nonlinear damping.  Some predictions based on the adiabatic approach have been compared with the results obtained from our direct numerical simulations. The expected exponential growth of the maximum of the solution was verified in our simulations, but  our simulations provide different power laws of the damping parameter $\delta$ in the exponential expression. Also, our measurements indicated that no mass is  dissipated in a single collapse event in the limit $\delta$ going to zero, different to the expected  finite amount of dissipated mass in the two-dimensional problem \cite{Fibich01}. Our findings were in agreement  with \cite{Fibich11,Fibich12}, showing the invalidity of the adiabatic approximation shortly after the arrest of the collapse. We provide a numerical evidence that it could be caused by the increasing influx of mass into the inner core of the solution.
	
	After the adiabatic regime, very close to the maximum of the solution, a quasi linear stage was observed as a consequence of the rapid defocusing process. Interestly, the validity of the universal profile (\ref{eq3_10}), after removing the interference oscillations caused by the extra mass, was verified in such quasi-linear regime. Thus, the post-blowup dynamics reduces mainly to an outward mass radiation  process. In these terms, and knowing that periodic boundary conditions allow that the mass radiated enter from the other side of the domain, we highlight the importance of using a kind of absorbing boundary conditions in order to prevent any unwanted interference with the dynamics in the interior of the domain. In addition, our observations suggest that in the limit $\delta$ going to zero, the critical mass is instantly radiated to infinity (in the case of the free-space domain) at the collapsing time.

\noindent \textbf{Acknowledgements.} J.M.E. thanks to IMPA for the financial support of his visit during the summer program of 2023. This work was supported by CNPq grants 308721/2021-7 and FAPERJ grant E-26/201.054/2022. 

\bibliography{references}
\bibliographystyle{plain} %{abbrv}

\end{document}